\documentclass[%
 reprint,
 amsmath,amssymb,
pre,
floatfix,
]{revtex4-1}

\usepackage{dcolumn}
\usepackage{bm}
\usepackage{amsmath}
\usepackage{arcs} 
\usepackage{amssymb,amsfonts,textcomp}
\usepackage[english]{babel}
\usepackage{xcolor}
\usepackage{array}
\usepackage{hhline}
\usepackage{graphicx}
\usepackage{lineno}
\usepackage[breaklinks=true]{hyperref}
\modulolinenumbers[5]
\hypersetup{colorlinks=true, linkcolor=blue, citecolor=blue, filecolor=blue, urlcolor=blue, pdftitle=, pdfauthor=, pdfsubject=, pdfkeywords=}
\usepackage{float}
\usepackage{breakcites}
\usepackage{upgreek}

\usepackage{float}
\usepackage{dblfloatfix}
\usepackage{fixltx2e}
\usepackage[section]{placeins}
\usepackage[center]{caption}
\usepackage{subcaption}

\usepackage{nomencl}

\newcolumntype{x}[1]{>{\centering\let\newline\\\arraybackslash}p{#1}}
\newcommand{\be}{\begin{equation}}
\newcommand{\ee}{\end{equation}}
\newcommand{\unit}[1]{$\,\mathrm{#1} $}
\definecolor{mygrey}{HTML}{FF0000}

\makeatletter
\newcommand\arraybslash{\let\\\@arraycr}
\makeatother

\usepackage{pdfcomment}

\begin{document}

\title{Generalized network modeling of capillary-dominated two-phase flow}

\author{Ali Q Raeini}
\author{Branko Bijeljic}
\author{Martin J Blunt}
\affiliation{Department of Earth Science and Engineering, Imperial College London, UK, SW7 2AZ}

\date{\today}

\begin{abstract}
We present a generalized network model for simulating capillary-dominated two-phase flow through porous media at the pore scale.      Three-dimensional images of the pore space are discretized using a generalized network -- described in a companion paper \citep{0Raeini2017a}  -- that comprises pores that are divided into smaller elements called half-throats and subsequently into corners.     Half-throats define the connectivity of the network at the coarsest level, connecting each pore to half-throats of its neighboring pores from their narrower ends, while corners define the connectivity of pore crevices.     The corners are discretized at different levels for accurate calculation of entry pressures, fluid volumes and flow conductivities that are obtained using direct simulation of flow on the underlying image.    This paper discusses the two-phase flow model that is used to compute the averaged flow properties of the generalized network, including relative permeability and capillary pressure.      
We validate the model using direct finite-volume two-phase flow simulations on synthetic geometries, and then present a comparison of the model predictions with a conventional pore-network model and experimental measurements of relative permeability in the literature. 

\end{abstract}



\maketitle

\section{ Introduction}

Modeling multiphase flow through porous media is important for understanding processes such as fluid flow in hydrocarbon reservoirs, contaminant transport, carbon storage in underground geological formations and fuel-cells.      Pore-scale modeling has been used to provide a link between the pore-scale fluid and rock properties to their macroscopic properties, such as relative permeabilities and capillary pressures.  Moreover, it can be considered as a complement to the experimental measurements of these parameters that are, in turn, used as input to field-scale models to predict the large scale behavior of flow \citep{0Dullien1992,0Sahimi1995,0Blunt2017Book}.

A variety of different methods have been used to investigate single and multiphase flow through porous media. These methods include molecular scale simulations studying fluid-rock interactions at nano meter scales to continuum-scale numerical methods such as lattice Boltzmann and finite volume, which can model two-phase flow directly on 3D images of the pore space, and pore-network models \citep{0Meakin2009}.      The following paragraphs present a brief review of the strengths and limitations of each of these methods.

Direct simulation, which solves the flow equations numerically while accounting for interfacial forces, is widely used to study two-phase flow at the pore scale.    Grid-based approaches, such as 
volume-of-fluid based interface capturing methods \citep{0Hirt1981, 0Gueyffier1999} 
 and diffuse-interface approximation of fluid-fluid boundaries using a density functional approach \citep{0Demianov2011, 0Koroteev2014}, have been applied to immiscible flows in porous media \citep{0Huang2005, 0Ferrari2013, 0Arrufat2014, 0Raeini2014b}.      
Particle-based methods such as lattice-Boltzmann have been used to compute absolute and relative permeabilities \citep{0Ferreol1995,0Martys1996}, capillary pressure \citep{0Pan2004,0Ahrenholz2008}, interfacial area \citep{0Porter2009}, and relative permeability for different rock types \citep{0Boek2010, 0Hao2010, 0Ramstad2010}.  
Another approach is to apply mesh-free methods such as smoothed particle hydrodynamics that, in addition to the computation of miscible flow and dissolution \citep{0Tartakovsky2009b}, have been employed to study two-phase flow through porous media \citep{0Sivanesapillai2015}.    
The advantage of direct methods is that the solid and fluid  interfacial boundaries can be modeled accurately.      Other features include the ability to study the effect of viscous forces, including the effects of flow rate and viscous coupling \citep{0Li2005, 0Ramstad2010, 0Raeini2014a}.      


Direct simulations, however, are computationally expensive.      Capturing layer flow through pore space crevices, for instance, requires a high resolution mesh.    Therefore, layer flow may not be captured accurately when using a coarse mesh that is usually required to make the simulations practical on larger images \citep{0Raeini2014b}.       More importantly, direct simulations may become impractical for simulating two-phase flow at low capillary numbers where small time-steps need to be used to resolve capillary waves and local instabilities (Haines jumps and snap-off) \citep{0Blunt2017Book}.      For most subsurface processes, flow occurs at very low capillary numbers and the flow domain can have heterogeneity at different scales.    Moreover, flow simulations may need to be run many times over a representative elementary volume that is several orders of magnitude larger than the grid resolution, to study the effect of input parameters  -- such as pore structure, contact angle, flow rates, fluid viscosities and initial conditions on the macroscopic properties  -- and to quantify the effect of uncertainties in these parameters.      A computationally efficient, and at the same time accurate, method to perform sensitivity studies to quantify the effect of these parameters is needed.      

The solution to this computational challenge is to use a multi-stage upscaling approach.      In the first stage, direct high-resolution simulations on smaller system sizes can be used to obtain the equations required to describe flow in individual pores and throats.      In a subsequent stage, these equations can be used to simulate flow through a coarser-scale network representation of the void space to obtain its macroscopic properties.

\subsection{Network modeling of two-phase flow}

Pioneered by \citet{0Fatt1956}, pore-network models have evolved as an important tool for studying flow through porous media.      They have been used to, for instance, investigate the effects of viscous \citep{0Lovoll2005, 0Li2005}, wettability and capillary forces, which control pore-scale configurations of fluids \citep{ 0Sahimi1995, 0Blunt2013}, discussed in more detail below.


\subsubsection{ Effect of wettability}

Rock adhesion forces, which give rise to a contact angle between a fluid-fluid interface and a solid wall, have a major impact on pore-scale displacement mechanisms.      Flow of the wetting phase through crevices of the void space (wetting layers), although slow, can lead to snap-off and hence trapping of non-wetting phases residing in centers of the void space.      This is in contrast to nanometer thick wetting films that are stabilized by molecular forces in strongly water-wet media, which can have a significant impact on apparent contact angle and capillary pressure \citep{0Tuller1999, 0Or1999} but their contribution to fluid conductivity is negligible \citep{0Blunt2002}.      Wetting layers, on the other hand provide a connected conduit for the wetting-phase flow down to low saturations \citep{0Lenormand1983, 0Blunt2002}.

Wetting layers can be modeled when using network elements with angular cross-sections, including fractal roughness models \citep{0Tsakiroglou2000}, grain boundary pore shapes \citep{0Mani1998, 0Man2001}, squares \citep{0Fenwick1998, 0Blunt1998} and triangles \citep{0Oren1998, 0Blunt2002}.      In most current network models, a shape factor, $G$, defined as the ratio of the cross-sectional area to the perimeter length squared, is used to assign the shape of pores and throats.   The shape of the cross-section -- a circle, square or scalene triangle -- is chosen such that it has the same shape factor as the 3D image of the porous medium \citep{0Oren1998, 0Patzek2001}.


The wettability of rock surfaces can change due to the adhesion of surface-active components of the oil to the solid surface \citep{0Buckley1998}.      The degree of wettability alteration depends on the composition of the oil and water, the mineralogy of the solid surface and the capillary pressure imposed during primary drainage \citep{0Buckley1998, 0Kovscek1993, 0Jadhunandan1995}.      The average wettability of a fluid rock system can be measured using core-flood experiments \citep{0Anderson1986b}.     More recently, contact angles have been measured directly in situ using micro-CT imaging \citep{0Andrew2014b,0Lv2017}.      Pore-network modeling has been used to link the pore-scale description of wettability to the bulk measurements as well as to study the trend in recovery with wettability \citep{0Jackson2003, 0Oren2003}.    Network models allow the incorporation of these in situ measurements of contact angle and its history dependence on a pore-by-pore basis.

\subsubsection{Effect of viscous forces}

Most quasi-static network models impose a single capillary pressure over the entire network.      This is used to define the fluid configuration in each element and the corresponding volumes of each phase.      Fluid interfaces, residing between pores and throats or in their crevices, change their configuration as the capillary pressure changes.    Filling of individual elements is assumed to take place over a much shorter time than the duration of the displacement process: this occurs in the form of Haines jumps, piston-like filling, layer collapse or snap-off events.      This is a valid assumption in most cases since typical capillary numbers ($C_A = {\mu U_D \over \sigma}$ where $\mu$ is the fluid viscosity, $U_D$ is the Darcy velocity  and $\sigma$ is the interfacial tension) in petroleum reservoirs are very low, in the range of $10^{-6}$ to $10^{-10}$. This represents a ratio of capillary to viscous pressure drop across a single pore of around 1000 or higher \citep{0Dullien1992}.      A single pore-filling event normally occurs in fractions of a second, as observed  using fast X-ray imaging or acoustic measurements \citep{0Berg2013,0Andrew2015,0DiCarlo2003}.      In contrast, it may take several days to years for a displacement process to be completed at a given location in a natural setting.      


The viscous pressure drop ($\Delta \varPhi$) can play a significant role when the flow rate is high (for example in near well-bore flow in hydrocarbon reservoirs), in near-miscible displacements with a low interfacial tension,  or when the length ($\Delta x$) of the representative elementary volume needed to obtain averaged properties becomes large.   The viscous pressure drop scales as $ \Delta \varPhi = U_D\mu\Delta x / K $, where $K$ is the effective rock permeability, while the capillary pressure, $P_c$, is not related to flow rate, $U_D$, nor system size, $\Delta x$.    Capillary pressure scales as $P_c = 2\sigma /r_c \sim \sigma \sqrt {\phi/K}$, where $\sigma$ is the interfacial tension, $r_c$ is the mean radius of interfacial curvature and $\phi$ is the porosity of the rock \citep{0Blunt2017Book}.    
When the ratio of viscous pressure drop to capillary pressure 
 is high, macroscopic flow properties, such as relative permeability, can be functions of flow rate, leading to a Darcy law where flow rate nonlinearly depends on the pressure gradient \citep{0Blunt2002}.  

To model two-phase flow when the viscous pressure cannot be ignored, the quasi-static assumption can be relaxed to consider the viscous pressure drop as a perturbation to the local capillary pressure along the length of the system.       A higher level of sophistication can be achieved by using a dynamic network-model that also incorporates the effect of changes in fluid volumes on the viscous pressure.      In these dynamic pore-models, the volume of each phase in each element should be updated using the flow rates from the computed pressure field.       Usually a very small time-step is required so that the configuration of fluid interfaces does not change significantly \citep{0Payatakes1982, 0Blunt1991, 0Aghaei2015, 0Regaieg2017}.     On the other hand, considering the viscous forces as a perturbation to the local capillary pressure will not cause a significant compromise in the computational efficiency.    In this approach the viscous pressure associated with rapid changes in the fluid configuration during a filling event is ignored, but the pressure gradient used in the computation of relative permeability is used as a perturbation to assign a non-uniform local capillary pressure across the system.     
We will adopt this perturbative approach in our model.

\subsubsection{Predictive capabilities of network models}

Pore-network models have reproduced particular experimental results of interest \citep{0Man2001,  0Tsakiroglou1999b, 
0Dixit2000, 0Fischer1999, 0Rajaram1997}.      This provides some assurance that the network models do represent flow and transport properties adequately.   There has been significant progress in the predictability of network models by constructing the element connectivity and shapes from the analysis of micro-CT images \citep{0Oren1998,0Sheppard2005,0Dong2009}.   However, \citet{0Bondino2013} showed that the algorithm used to generate and parametrize the network model using current network extraction algorithms can have a significant impact on the predicted macroscopic properties.  
In other words, the description of the void space in current network models -- using pore and throat lengths, shape factors, radii and volumes -- is inadequate for the reliable prediction of two-phase flow using micro-CT images of porous media.    Therefore, improving the predictability of pore-network extraction and flow modeling remains an active research topic  \citep{0Idowu2013,0Miao2017,0Gostick2017,0Ruspini2017}.    The problem is that the conventional network model parameters cannot be independently verified and therefore the model primarily relies on calibration to predict experimental measurements.  Therefore, there is little confidence that a network calibrated using a small set of benchmark experiments can predict the properties of different rock types or different displacement scenarios reliably.        To avoid this problem, the  geometry of individual pores in the network model should be as close as possible to the real system.

We have presented a generalized network representation of the void space that eliminates the need for intermediate parameters, specifically shape-factors and pore and throat lengths, used to describe the network elements \citep{0Raeini2017a}.      Instead, the parameters required for network modeling are extracted using direct single-phase simulation and a medial-axis analysis of the void space.       Each pore is subdivided into half-throats and further into corners.       The parameters approximating each corner -- corner angle, volume and conductivity -- are directly extracted from the underlying micro-CT image at different discretization levels and exported as tabulated data to the network flow simulator.

The emphasis of this paper is on the formulations used to describe capillary dominated two-phase flow through the generalized network representation of the void space, and on the prediction of multiphase flow properties -- capillary pressure and relative permeability.   
We present the algorithms used to track the fluid-fluid interfaces and incorporate the effect of contact angle that describes the wettability of the fluid-rock system.  Gravity and viscous forces are treated as a perturbation to compute the local capillary pressures throughout the network.

\section{Generalized network flow modeling}

The network model can be viewed as an upscaled alternative to direct two-phase flow models.    
It is comprised of three main components: (a) parametrization and tracking of fluid interfaces in each element as the local capillary pressure ($P_c$) changes during a displacement cycle, (b) tracking fluid phase distribution and connectivity, and (c) computation of fluid-phase saturations and conductivities for individual elements and for the whole system.

The flow simulations are designed to represent typical core-flood experiments. Initially, the whole void space is assumed to be filled with water, and a second fluid, oil, is injected from one side of the network to obtain the fluid configurations at the end of a primary oil-invasion simulation. 
The simulations are continued for a second cycle, water flooding, in which the water pressure is increased to replace the oil. The flow simulations can be continued for a third cycle by injecting oil, to obtain, among other properties, wettability indices.

The following sections present the details of our network flow model, which involves tracking the fluid configurations within the void space (see Figure \ref{fig_illustrate}) and upscaling their flow properties.
An overview of how the void space is represented using half-throat corners is discussed in Section \ref{sec_elemInit}.
Fluid configurations in a corner are described using their interfaces with other fluids in the corner and their connectivity to fluids in the surrounding pores and throats; this is explained in  Section \ref{sec_connectivity}.   Section \ref{sec_layerDetailed} presents the details of how fluid interfaces are represented and tracked.
In Section \ref{sec_Pes}, we describe the computation of entry pressures required for fluid interfaces to change configuration.  These include filling of centers of pores and throats, and growth and collapse of water and oil layers in edges of void space corners.  
In Section  \ref{sec_filling}, we discuss how displacement events (changes in fluid configurations)  affect fluid phase connectivity: forming disconnected phases, newly connected phases (coalescence) and exposing new invasion paths for subsequent displacement events.    Finally, in Section \ref{sec_Kr}, we discuss the assignment of fluid volumes and conductivities that are then used in the computation of the fluid saturations and relative permeabilities of the network.

\subsection{ Description of the network elements} \label{sec_elemInit}

\begin{figure}[H]
\centering 
 \includegraphics[width=0.42\textwidth]{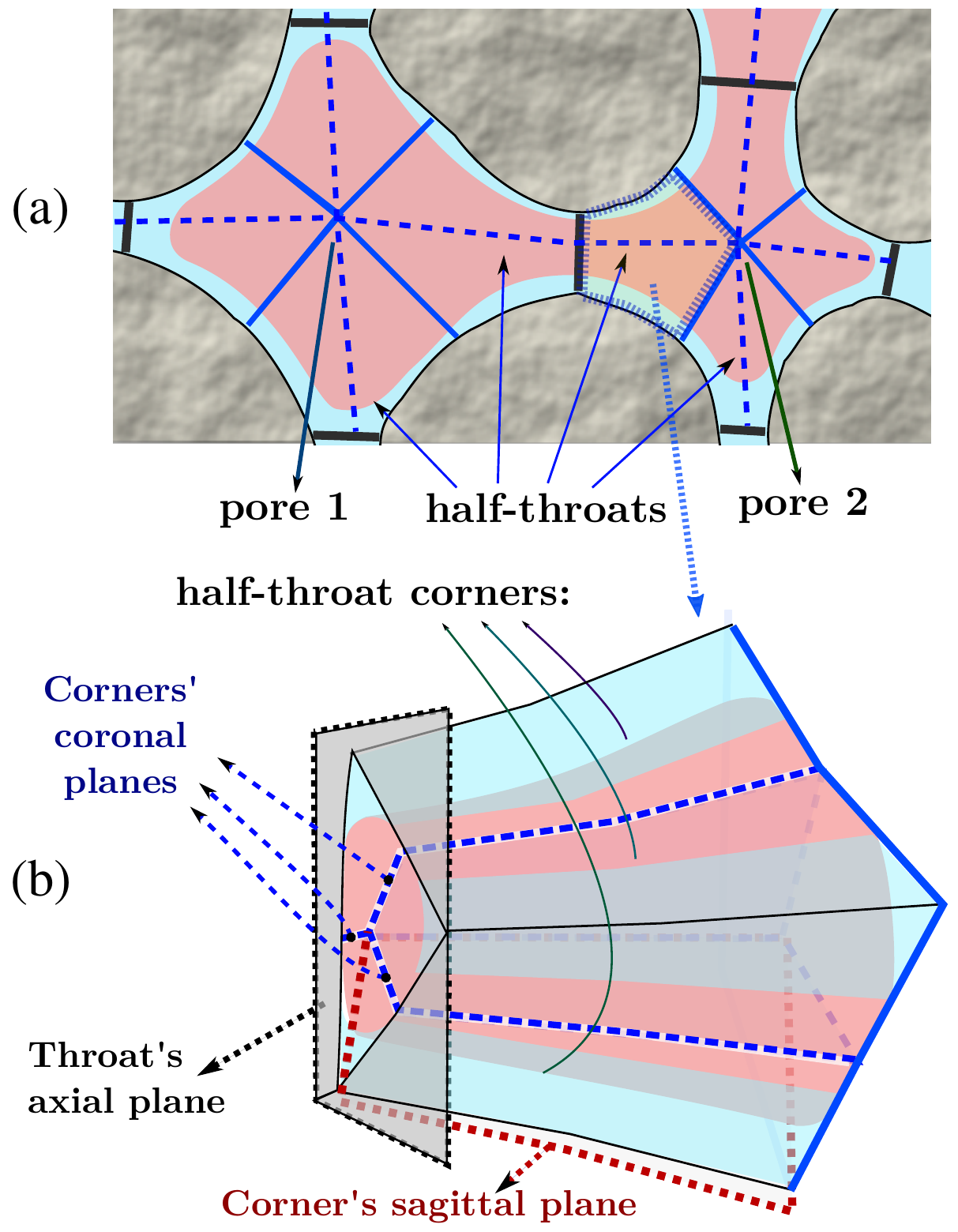}
 \caption{An illustration of two fluids (highlighted in red and light blue) occupying a void space that is discretized into (a) pores and half-throats and further into (b) corners. 
The thick solid black lines show the boundaries between pores (the throat surfaces).  The solid blue lines show the partitioning of the pore space into half-throats. The dashed blue lines show the boundary between the corners of each half-throat. 
} \label{fig_illustrate} 
\end{figure}

Figure \ref{fig_illustrate} shows how the void space is divided into pores, half-throats and their corners.
We use a watershed segmentation of the distance map (the distance of any point in the void space to the nearest solid) of the underlying image to divide the void into pore regions bounded by throat surfaces \citep{0Raeini2017a}.  We further divide these pore regions into half-throats and their associated corners.  Every point in the void space is uniquely assigned to a pore, a throat and a half-throat corner.   We refer to the half-throat corners simply as corners in what follows.

\begin{figure}[H]
\centering 
 \includegraphics[width=0.38\textwidth]{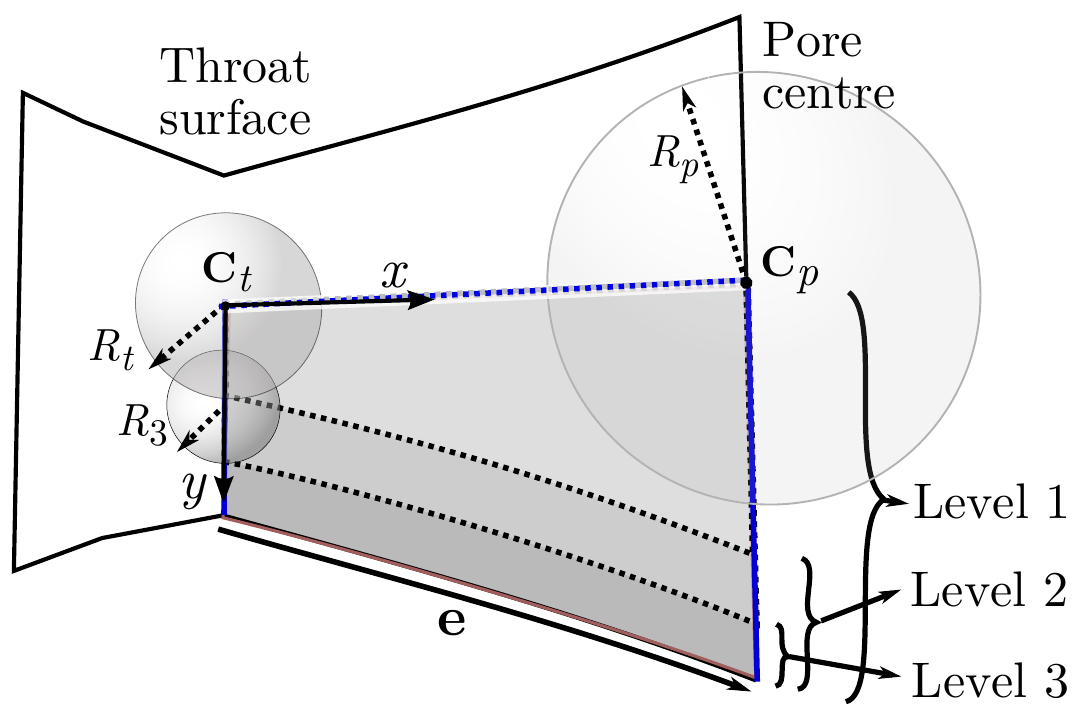}
 \caption{An illustration of different corner discretization levels, $i=1$-$3$ in the corner's sagittal plane. Each level consists of the void space outside maximal inscribed spheres with radius $R_i=\,R_p,~R_t$ and $0.7R_t$, respectively. $x$ and $y$ are the local coordinates of the corner and $\bf e$ is the edge vector.} \label{fig_levels}
\end{figure}

The corners are parametrized at different discretization levels, $i=1$-$3$, which are obtained during network extraction. Each discretization level contains the corner void space outside maximal-spheres with radius larger than $R_i$. In this paper, we choose $R_1=R_p$, $R_2=R_t$, $R_3=0.7R_t$, see Figure \ref{fig_levels}.

We define a local coordinate, ($x,y$), for each corner (Figure \ref{fig_levels}): $x$ represents the distance from the throat surface and $y$ represents the distance from the throat line in the corner's sagittal plane. The $x$ axis is aligned with the throat line and the $y$ axis is aligned with the medial axis of the corner. The corner edge is defined as the line where the two sides of the corner meet and the vector along the corner edge that connects the throat surface to the corner's axial cross-section at the pore center is called the edge vector,  $\bf e$.

The inscribed radii and cross-sectional areas are used to compute discretization level depths ($H_i$) measured along the corner sides, and half angles ($\gamma_i $), as  illustrated in Figure \ref{fig_Hc} and discussed in Appendix \ref{sec_elemInitAppendix}.

\begin{figure}[H]
\centering 
 \includegraphics[width=0.44\textwidth]{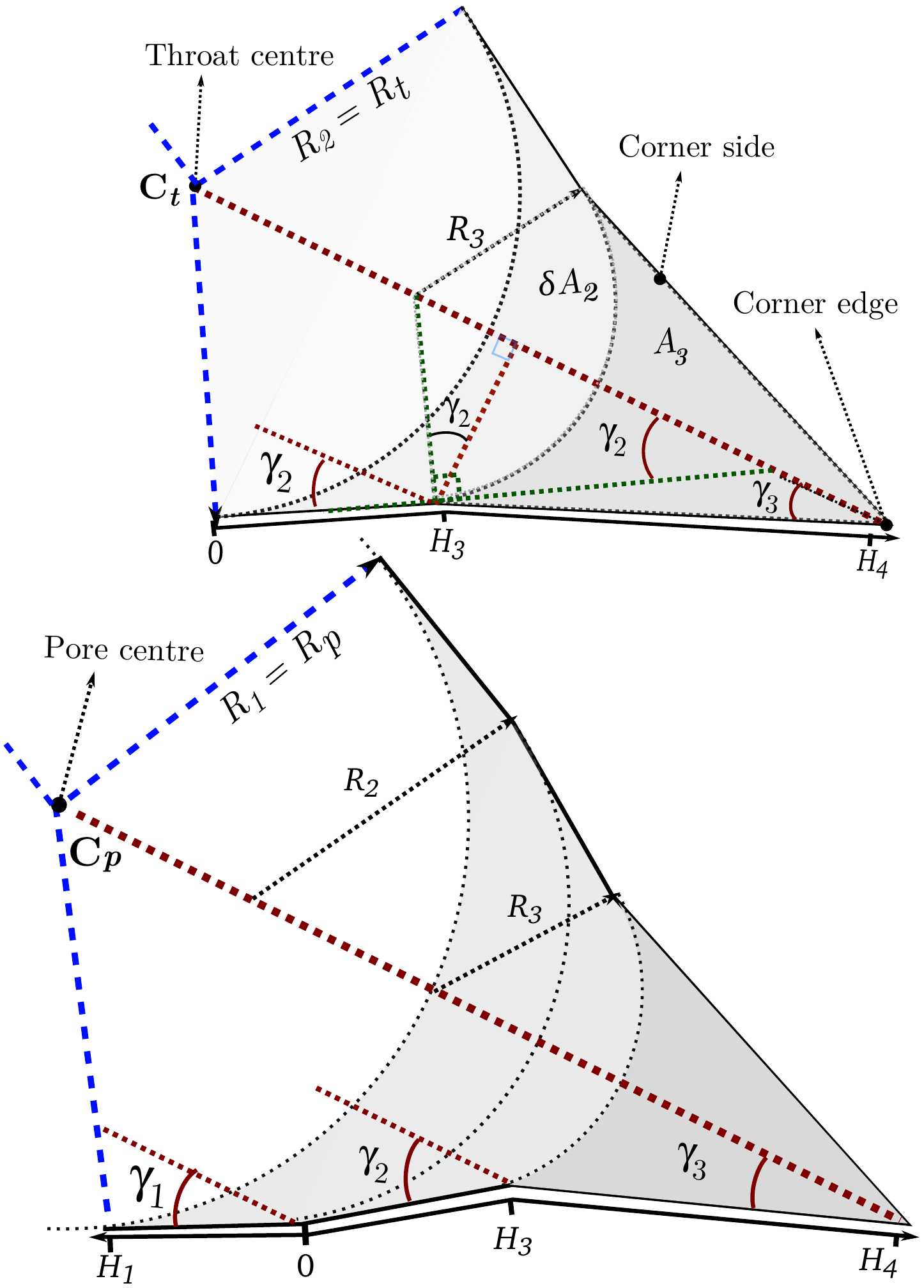}
 \caption{Cross-sections of a corner in its axial planes passing through the throat surface (top), and through the pore center (bottom). The dashed blue lines show the boundary between the corner and other corners present in the half-throat that are not shown in this figure for simplicity. The dotted circular arcs separate the corner discretization levels. The dotted thick red line shows the corner \emph{center line} (medial axis) and the solid black lines show the two \emph{side}s of the corner that meet at the corner \emph{edge}.} \label{fig_Hc}
\end{figure}

The extracted corner parameters for each discretization level consists of the inscribed radius ($R_i$), cross-sectional area in the corners axial plane ($A_i$), volume ($V_i$), and flow ($g_q$) and electrical ($g_e$) conductivities.   These properties come from the analysis of an underlying pore-space image \citep{0Raeini2017a}.

$R_i$, $H_i$ and $\gamma_i$ define the geometry of the corners and are used to semi-analytically track the location of fluid interfaces through the pore space as the local capillary pressure changes.  $V_i$ and $g_i$, on the hand, define the volumes and conductivities of the corners, which are used to compute fluid saturations and relative permeabilities of the network.

\subsection{Connectivity and fluid configurations} \label{sec_connectivity}

\begin{figure}[H]
\centering 
 \includegraphics[width=0.49\textwidth]{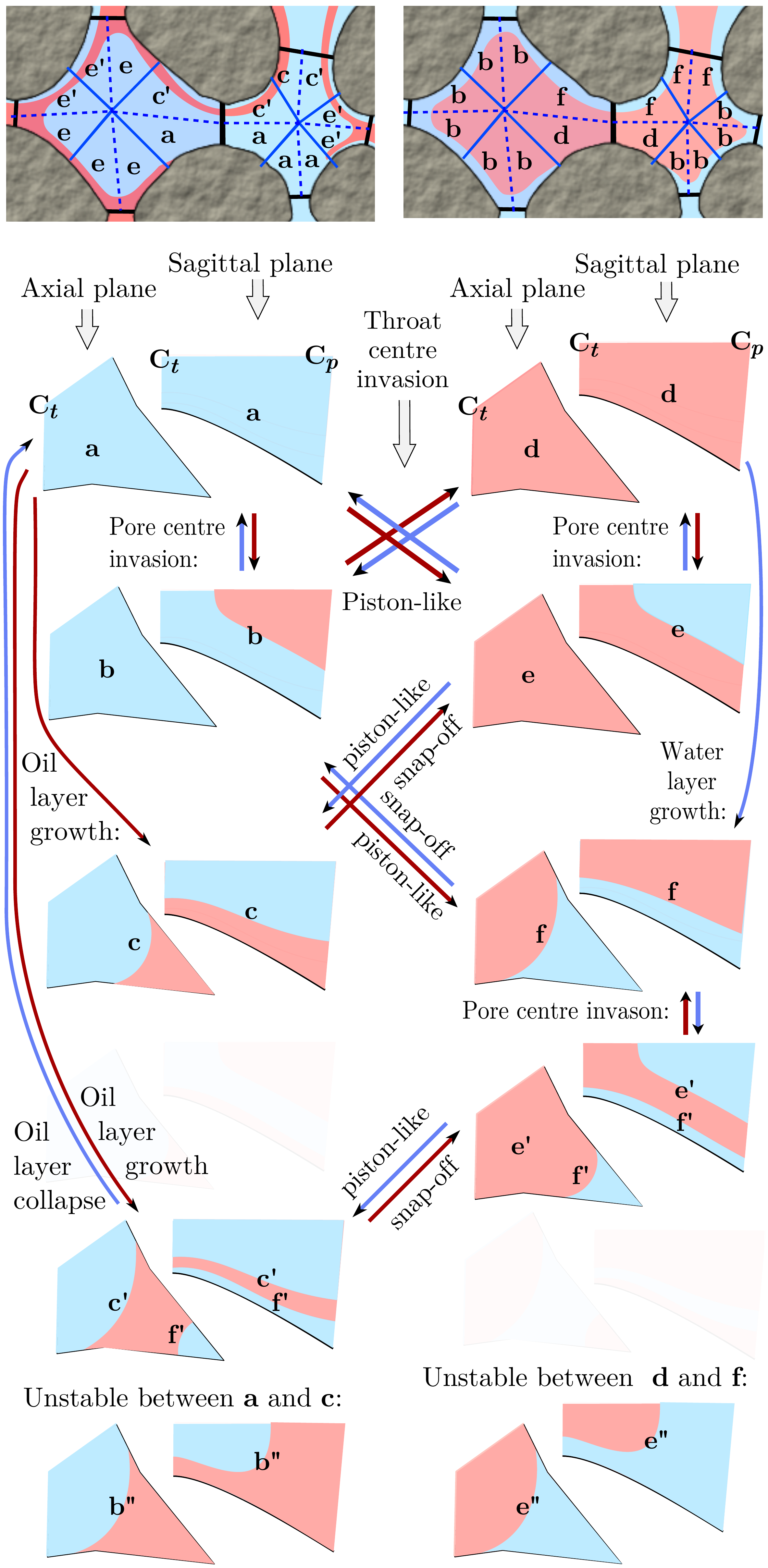}
 \caption{An illustration of fluid phase connectivity (top row), and different fluid configurations and displacement events. Red  areas represent the oil phase and water is shown in light blue. Dark red arrows show invasion by oil and blue arrows show invasion by water. The fluid interfaces are classified into two groups:  b, b$''$, e, e$'$ and e$''$ are called piston-like configurations while  c, c$'$, f and f$'$  are called layers.   } \label{fig_fluid_config}
\end{figure}

How we define fluid configurations and assign their connectivity is shown schematically in Figure \ref{fig_fluid_config}.
The fluid configurations during a flow simulation are modeled by considering four \emph{flow path}s for each corner: (1) in the throat center, (2) in the pore center, (3) in the corner edge and (4) sandwiched between the corner edge and corner center. Different fluid configurations are then constructed by marking each flow path as filled by oil ($\alpha=o$) or water ($\alpha=w$), subject to the following rules. (i) Corners share the same throat center, the fluid occupying a throat center occupies all its corner centers. Similarly, (ii) the fluid in the pore center is shared between all of the pore corners.  (iii) Only oil layers are considered to be sandwiched between water in the edge and in the center of a corner.

If two \emph{adjacent} flow paths are occupied by different fluids,  a fluid-fluid interface is allocated between them. A flow path can grow or shrink in size if the interface separating it from other fluids in adjacent paths move between different corner discretization levels, or along the same discretization level.

Corner edges  and corner centers are considered as adjacent flow paths.  However, if there is an oil layer sandwiched between a water layer and water in the center, the two water phases are considered as disconnected from each other.   The interface between a pore center and a throat center is called a piston-like interface. An interface separating two fluid layers, or a fluid layer and a fluid in the center, is called a layer interface or simply a layer.  Sections \ref{sec_layerDetailed} and \ref{sec_pistonDetailed} discuss how these two interface configurations are represented.

Another factor controlling a fluid configuration is its connectivity with fluids in the surrounding pores.
At the throat surface, the corners on either side, belonging to neighboring pores, are connected by definition.  At a pore center, each corner does not necessarily connect to every other corner that is associated with the pore's half-throats, discussed next.  This is where our approach differs from conventional network models (eg. \citep{0Valvatne2004, 0Patzek2001, 0Oren2003}) where it is assumed that at a pore all the corners are connected to each other.

We only allow a corner to be connected to one or two other corners belonging to different throats, based on their proximity. 
We first find the adjacent throats, up to two throats, whose throat lines have the smallest angle with the corner's $y$ axis (see Figure \ref{fig_levels}). Then, among corners of each of these throats, we find the adjacent corner that has the smallest angle between its $y$ axis and the corner's $y$ axis. If the angle is less than 60 degrees, the two corners are assumed to be connected to each other and are called adjacent corners.  
We only allow connections to corners in different throats: fluids in the corner edges of a throat are not directly connected to each other. 

The water layers residing in each corner are considered to be connected to the water inside their adjacent corners. Similarly, oil layers are considered to be connected to oil phases in their adjacent corners. These adjacent fluids can themselves be in layer, piston-like or single-phase configurations.

During network extraction, throats that are connected to the left and right sides of the image boundary are considered as boundary throats.  In this paper, for the sake of simplicity, we call the left side boundary as the inlet and the right side as the outlet.  The connectivity of each fluid to outlet and inlet throats is obtained using a graph search through the adjacent flow paths containing the fluid.   Fluids that are not connected to the outlet throats are considered trapped and do not change configuration. 
Similarly, the invading fluid connectivity, which is injected from the inlet throats, affects the order that the new interface configurations are formed, as discussed in Section \ref{sec_filling}.  In addition, fluid connectivities affect the computation of their conductivity (discussed in Section \ref{sec_Kr}), and computation of interface curvatures and entry pressures for different displacement events, which are discussed next.

\subsection{Layer configurations } \label{sec_layerDetailed}

\begin{figure}[H]
\centering 
 \includegraphics[width=0.47\textwidth]{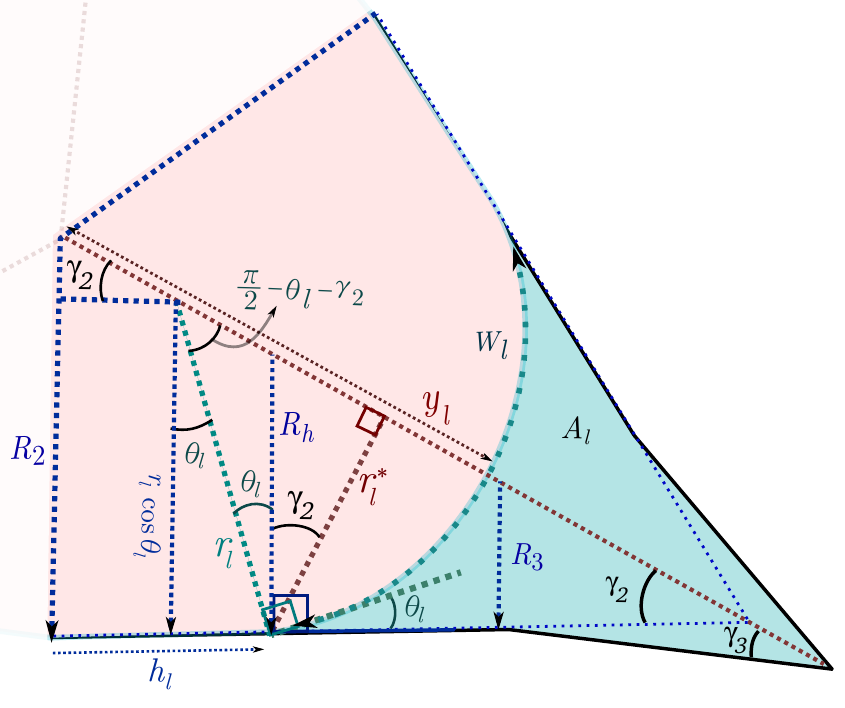}
 \caption{ An illustration of the parameters used to describe the location, ${h_l}$ (Eqs.\,\ref{eq_hl}-\ref{eq_pc}), of a layer (darker blue color) located at the level 2 of a corner at the throat surface, and its relationship with contact angle ($\theta_l$), radius of interface curvature in the throat's axial plane ($r_l$, Eq.\,\ref{eq_rl}),  and the interface distance ($y_l$, Eq.\,\ref{eq_hlc}) from the corner center along its medial axis. } \label{fig_Hl}
\end{figure}

The location of a layer interface ($h_l$) -- that can be oil or water or a sandwiched oil layer -- in a corner is defined as the distance of its interface contact line from the center of the pore or throat, which is computed along the side (solid wall) of the corner as illustrated in Figure \ref{fig_Hl}.   The sandwiched oil layers, which have two interfaces (see Figure \ref{fig_fluid_config}), are described using their interface with the water in the center.   The location of the other interface, between the oil layer and the water in the edge, is referred to as the water layer interface.   
A layer, if present, is tracked using its interface location, which at the throat surface can reside at discretization levels 2 and 3.  Near the pore center, however, layer interfaces can reside additionally at level 1 -- this can happen in a piston-like configuration as discussed in Section \ref{sec_pistonDetailed} or due to the variations in a layer interface curvature in its sagittal plane (Section \ref{sec_sagittal}).

To uniquely describe a layer configuration, in addition to $h_l$, we need to know the contact angle, $\theta_l$, between its interface and the solid wall.
For simplicity, we define the layer contact angle as the angle between the layer interface and solid wall that is measured through the layer ($l$) that is closer to the corner edge. This definition is not the same as the conventional definition as an angle  ($\theta_w$) that is always measured through the denser phase ($\theta_w =  \theta_l $ for water layer interfaces and $\theta_w = \pi-\theta_l $ for oil layer interfaces).    The contact angle is history-dependent and can vary between a receding and an advancing contact angle ($\theta_r$ and $\theta_a$ respectively) that are input into the network flow model.

The location, $h_l$, of a layer interface (Figure \ref{fig_Hl}), residing in a corner discretization level, $i$, (Figure \ref{fig_Hc}) for a given contact angle, $\theta_l$, is:
\be \label{eq_hl}  h_l = H_i + \frac{R_i\cos {\gamma_i} - r_l \cos (\theta_l +\gamma_i) }{ \sin {\gamma_i} },~~~~ i=1\text{-}3\ee
where $H_i$ (Eq.\,\ref{eq_Hp}) is the corner level depth and $\gamma_i$ (Eq.\,\ref{eq_hAng}) is its half angle.  $r_l$ is the interface radius of curvature in the corner's axial plane.   $r_l$ can be obtained from the local capillary pressure $P_c$:
\be \label{eq_pcSignlayer}
P_c= 
\begin{cases}
\sigma\kappa_l & \alpha_t=w \\
-\sigma\kappa_l & \alpha_t=o 
\end{cases},
\ee
\be \label{eq_pc}  \kappa_l=\frac{1}{r_l} + \frac{1}{r_s}.  \ee
$\kappa_l$ is the interface total curvature and $r_s$ is the radius of the layer curvature in the sagittal plane of the corner, discussed in Section \ref{sec_sagittal} in detail.

To allow a unique assignment of  the contact angle  and interface location, we need to record the initial interface location as the layer configuration forms and track it as the simulation progresses.  We initially set $h_l=H_4-\epsilon$ (near the corner edge, see Figure \ref{fig_Hc})  if the interface forms due to layer growth, and $h_l=H_1+\epsilon$ if the interface is left behind during a piston-like invasion of the corner center (see Figure \ref{fig_fluid_config}). $\epsilon$ is a small number, set to $10^{-9}$\,m in this paper.

\begin{figure}[H]
\centering 
 \includegraphics[width=0.40\textwidth]{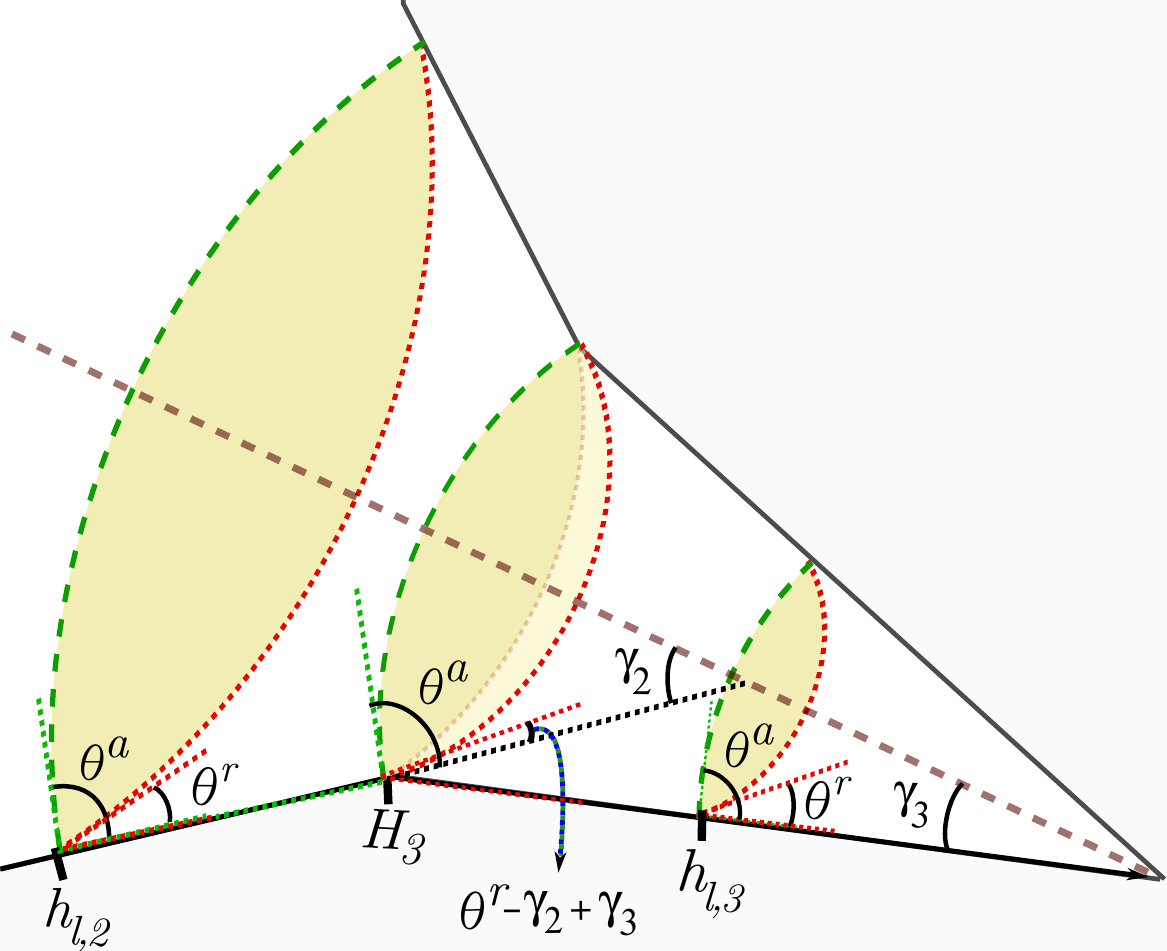}
 \caption{An illustration of contact angle hysteresis in a segment of a corner cross-section showing levels 2 and 3. The dotted red lines show the interface at different locations ($h_{l,2}$, $H_3$ and $h_{l,3}$) as it recedes from discretization level 2 to level 3. The dashed green lines show the interface advancing from level 3 to level 2. Yellow areas show where the interface can swing while being pinned (i.e. its contact line with solid remains fixed at $h_l$). } \label{fig_CAHist}
\end{figure}

We then use a multi-stage computation to find a unique solution for $\theta_l$ and $h_l$ from Eq.\,\ref{eq_hl} for a given capillary pressure or interface curvature.
First we find the discretization level, $i$, in which the interface has been previously residing (see Figure \ref{fig_CAHist}): 
\be \label{eq_level}
i\in \{1,2,3\}: ~~~H_i<h_l \leq H_{i+1}.
\ee
At the throat surface, however, $i$ can be 2 or 3 only.

If Eq.\,\ref{eq_hl}, with this $i$ and $h_l$ fixed to its previous location gives a $\theta_l$  in the range [$\theta_r$, $\theta_a$], it is assumed that the location of the interface remains pinned at its original position and the computed $\theta_l$ is accepted as the so-called hinging contact angle. Otherwise, the contact angle is fixed to $\theta_r$ when the layer pressure decreases, or to $\theta_a$ when the layer pressure increases, and Eq.\,\ref{eq_hl} is used to compute the new interface location. 

If the computed interface location using Eq.\,\ref{eq_hl} does not correspond to its previous level based on Eq.\,\ref{eq_level}, the interface will move to a new discretization level that satisfies Eq.\,\ref{eq_level}, or will be pinned at their boundary. If (a) $\gamma_2<\gamma_3$, the interface is assumed to jump to the new corner level and  Eq.\,\ref{eq_hl} is invoked with the new discretization level corner angle to recompute the new interface location.  However, if (b) $\gamma_2>\gamma_3$, the interface gets pinned at $h_l=H_3$ before moving to the next discretization level.   In this case, we first check if the interface has a stable position at $h_l=H_3$ assuming $\gamma=\gamma_2$ with a contact angle between $[\theta_r-\gamma_2+\gamma_3$, $\theta_a]$.   If Eq.\,\ref{eq_level} has a solution in this range, the interface will be pinned at $h_l=H_3$, otherwise it will move to the new corner level and  Eq.\,\ref{eq_hl} is invoked with the new level corner angle to recompute $h_l$.

The same algorithm is used in the piston-like interface to find the location of its tailing layers that reside at the pore center but not at the throat surface (see Figure \ref{fig_Hl_pl}), when they move between different discretization levels $i=1$-$3$.

$h_l$ and the contact angle, $\theta_l $, can be used to obtain other interface parameters (see Figure \ref{fig_Hl}), including the width of the layer interface, ${W_l}$, and the distance of the interface center from the throat center, $y_l$:
\be  \label{eq_rl} r_l = \frac{R_i\cos \gamma_i- (h_l-H_i) \sin \gamma_i}{\cos (\theta_l +\gamma_i)},  \ee
\be \label{eq_Ll}  W_l = (\pi - 2\gamma -2\theta_l) r_l,  \ee
\be \label{eq_hlc} y_l = \frac{ {R_{ci}-r_l}cos({\theta_l })}{\sin \gamma } + r_l.  \ee

The corner inscribed radius, $R_h$, the radius of largest inscribed sphere touching the solid wall at $h_l$ (see Figure \ref{fig_Hl}), is assumed to vary linearly between the corner discretization levels $i$ and $i+1$:

\be \label{eq_R_h} {R_h}=R_i+\frac{\updelta R_i}{\updelta H_i}(h_l-H_i)   \ee
\noindent where $\updelta$ is the difference operator: $\updelta R =R_i = R_{i+1}$ and $\updelta H = H_i-H_{i+1}$. To obtain the layer cross-sectional area, $A_l$, we first compute the cross-sectional area for a hypothetical layer located at $h_l$ but with a contact angle of zero, $A_h$,  and then add the effect of contact angle, $A_\theta$:
\be \label{eq_A_l} A_l = A_h+ A_\theta, \ee
\be \label{eq_A_h} A_h=A_i-(R_i^2-R_h^2)(\frac{1}{\tan \gamma_i}-\frac{\pi}{2}+\gamma_i),  \ee
\be \label{eq_A_CA} A_\theta = r_l^{*2} (  \frac{\frac{\pi}{2} -\gamma_i }{\cos^2\gamma_i}
							+  \tan \theta^* - \tan\gamma_i   - \frac{\frac{\pi}{2} -\theta^*}{\cos^2 \theta^* }), \ee  
\noindent where $\theta^*=\theta+\gamma_i$ and $r_l^*=r_l\cos(\theta^*)$.
Note that for the case of oil layers sandwiched  between water in the edge and in the center, $A_l$, obtained using Eq.\,\ref{eq_A_l}, includes the cross-sectional area of the oil as well as the water in the corner.

\subsection{Layer description in a corner's sagittal plane} \label{sec_sagittal}

\begin{figure}[H]
\centering 
 \includegraphics[width=0.39\textwidth]{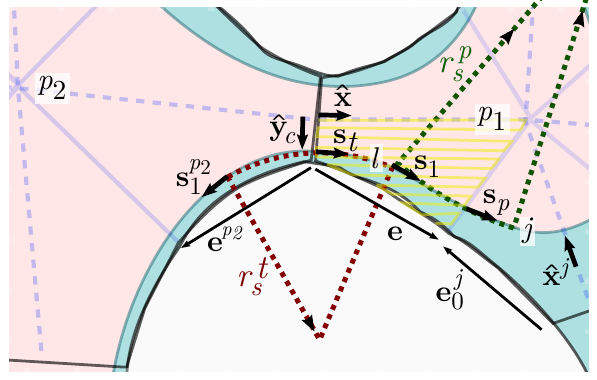}
\caption{A schematic representation of the parameters used to estimate a layer (blue areas) radius of curvature ($r_s$ -- Eqs.~\ref{eq_rlSag_t} and \ref{eq_rlSag_p}) in the sagittal plane of a corner (highlighted using horizontal yellow stripes). The sign of $r_s^t$ is negative while  $r_s^p$ is positive in this case. $j$ is the adjacent corner in the same pore and $p_2$ is the adjacent pore. $ \bf e$ is the tangent vector to the corner edge and $\bf s$ is the unit vector tangent to the layer interface in the corner's sagittal plane. $\hat{\bf x}$ and $\hat{\bf y}$ are the unit vectors along the corner local coordinates, $x$ and $y$. } \label {fig_sagital} 
\end{figure}

The layer interface radius of curvature in a corner's sagittal plane ($r_s$) is needed in Eq.\,\ref{eq_pc} to relate the radius of curvature in the corner's axial plane ($r_l$) to the local capillary pressure.  We estimate it from the tangent vectors, ${\bf s}_1$ (Eq.\,\ref{eq_s_1}), to the interface in the corners mid-sagittal plane at a distance $x=x_1=x_p/2$ from the throat surface, see Figure \ref{fig_sagital}.

At the throat center, $r_s$ is estimated from:
\be \label{eq_rlSag_t} \frac{1}{r_s^t} = -  2\frac { {\bf s}^{p1}_{1}+ {\bf s}^{p2}_{1} } { |{\bf e}_{~}^{p1}+ {\bf e}_{~}^{p2}|} \cdot {\bf \hat{y}}_c.  \ee
Similarly, $r_s$ at the pore center is estimated from:
\be \label{eq_rlSag_p} \frac{1}{r_s^p} = -  2\frac { {\bf s}_{1}+ {\bf s}^j_{1} } { |{\bf e}_{\,}+ {\bf e}_{\,}^j|} \cdot \frac{{\bf \hat{y}}_c+{\bf \hat{y}}_c^j}{|{\bf \hat{y}}_c+{\bf \hat{y}}_c^j|}  \ee

\subsection{Piston-like configuration } \label{sec_pistonDetailed}

\begin{figure}[H]
\centering 
 \includegraphics[width=0.35\textwidth]{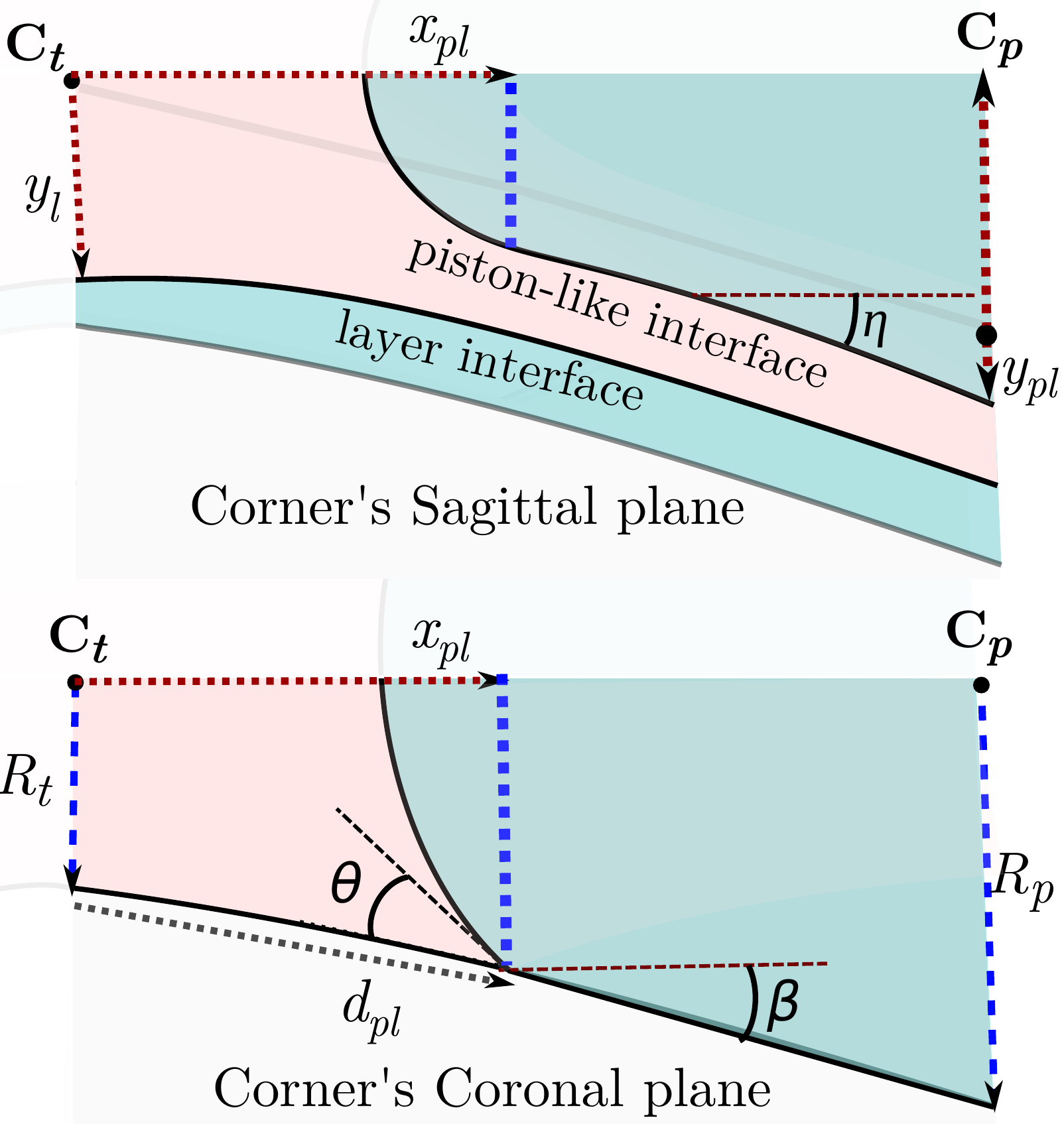}
 \caption{An illustration of a piston-like interface in a corner's (top) sagittal plane, and (bottom) coronal plane -- the plane perpendicular to each of the corner sides and passing through the pore and throat centers; see Figure \ref{fig_illustrate}. The interface curvature is obtained by writing a force balance on the interface to the left side of the thick dotted blue line, in the $x$ direction (Eq.\,\ref{eq_pl_k}).} \label{fig_Hl_pl} 
\end{figure}

A piston-like configuration,  Figure \ref{fig_Hl_pl}, refers to a fluid-fluid interface separating a pore center from a throat center; these interfaces are often called terminal menisci, since they block the center of the pore space \citep{0Blunt2017Book}.

The curvature of a piston-like interface, $\kappa_{pl}$, is controlled by the interface contact angle with the solid walls and the configuration of its tailing layers residing in the corner edges.    It is obtained by writing a force balance equation on the interface:
\be \label{eq_pl_k} \kappa_{pl} = \frac{ \sum {2 h_l \cos (\theta + \beta) + W_l {\bf s}_l \cdot {\bf \hat{x}}  }}{A^t_x - \sum A_l}\ee
where $A^t_x$ is the throat total cross-sectional area at a distance $x$ from the throat surface, Eq.\,\ref{eq_A_x}. The summations ($\sum$) are performed over all the throat corners, $c=1$-$n_c$, where $n_c$ is the total number of corners in the throat.  $l$ stands for the layer in the corner, $c$, that is adjacent to the fluid in the center.  If the layer does not exist in the corner, its area ($A_l$) and arc length $W_l$ are set to zero and $h_l=H_4$.

$\beta$ is the angle between the corner side plane and the throat line -- the line connecting the throat center to the pore center, see Figure \ref{fig_Hl_pl}, computed using Eq.\,\ref{eq_beta}.  

We compute the piston-like curvature at three points along the throat line, ${x}$, at the throat center (${x}_0=0$), at the pore center (${x}_2={x}_p$), and in between at ${x}_1={x}_p/2$. The interface curvature is assumed to vary linearly from $x=x_t$ to $x_1$, and from $x=x_1$ to $x_p$.
\be \label{eq_pl_x}
\kappa_{pl}(x)=\kappa_{pl}^i+(\kappa_{pl}^{i+1}-\kappa_{pl}^i)\frac{x-x_i}{x_{i+1}-x_i}, ~~~ i=0,1
\ee
%

The sign of $\kappa_{pl}$ is considered positive when the interface is curved toward the pore center, hence:
\be \label{eq_pl_sign}
P_e(x)= 
\begin{cases}
\sigma\kappa_{pl} & \alpha_p=\alpha_{inv} \\
-\sigma\kappa_{pl} & \alpha_p\neq\alpha_{inv} 
\end{cases},
\ee

A location is assigned to piston-like interfaces, defined as the distance of the interface contact-lines from the throat surface, $x_{pl}$. Similar to the layer interface location $h_l$, $x_{pl}$ is assigned as soon as a pore or a throat center is filled by a fluid, forming the piston-like configuration, and tracked as the simulation progresses.
The interface is assumed to remain pinned (fixed $x_{pl}$) as long as the invading phase pressure is below $P_e(x_{pl})$ (Eq.\,\ref{eq_pl_sign}). Once the invading phase pressure surpasses $P_e(x_{pl})$, the interface location is updated by solving Eqs.\,\ref{eq_pl_x} and \ref{eq_pl_sign} for $x_{pl}$ (replacing $x$ with $x_{pl}$). 

The computed interface curvature and location are used to assign the pore and throat entry pressures and to compute the fluid volumes and conductivities, which are described in the following sections in more detail.


\subsection{ Computation of entry pressures} \label{sec_Pes}

As discussed in Section \ref{sec_connectivity}, we consider four  \emph{flow path}s for each corner: in the throat center, in the pore center, in the corner edge and sandwiched between corner edge and corner center. A displacement is a change in the occupancy of a flow path and involves a reconfiguration of fluid interfaces.   The entry or threshold pressure, $P_e$, is the pressure required to overcome the interfacial force as the interface passes through the flow path during this reconfiguration.  $P_e$ is defined as the maximum relative pressure, the invading phase (subscript $inv$) pressure minus the receding phase pressure, encountered during each displacement.

\be
P_e= 
\begin{cases}
 ~~P_{c,max}, & \alpha_{inv}=o \\
 - P_{c,min}, & \alpha_{inv}=w 
\end{cases}.
\ee


\subsubsection{ Throat center entry pressure}

For throats there are (a) one entry-path from each of the two neighboring pores (piston-like displacement), and (b) one from each corner (snap-off displacement), if they contain the invading phase and are connected to the inlet.

{\bigskip ~~~ a) Piston-like displacement \smallskip}

The throat entry pressure by piston-like invasion is computed by solving Eqs.~\ref{eq_pl_k} and \ref{eq_pl_sign} at the throat center, by iteratively changing the interface curvature, $\kappa_{pl}$, using the Newton-Raphson method. 

{\bigskip  ~~~ b) Snap-off entry pressure \smallskip}

The throat snap-off pressure is approximated from the lowest interface curvature (measured toward the center/receding phase) that the interface experiences as it moves from its initial location toward the throat center meeting the interface of other throat corners.  It is approximated as the minimum of the $k_l$, Eq.\,\ref{eq_pc}, obtained from (a) solutions of Eq.\,\ref{eq_rl} with $h_{l}$ varying from its initial location to $h_{l}=0$ and (b) the solution of Eq.\,\ref{eq_hlc} for $ y_l=0$.


Note that both $h_l$ and $y_l$ are history dependent due to contact line pinning (see Figure \ref{fig_CAHist}). The effect of contact line pinning is considered in the calculation of $h_l$ and $y_l$, in all the equations that require the computation of interface location using Eqs.\,\ref{eq_hl}-\ref{eq_level}. 

\subsubsection{ Pore filling pressure} \label{sec_poreFilling}

A pore center can be invaded from any of its throats that contain the invading phase.      The pore center entry pressure, $P_e^p$, is the \emph{smallest} of all the threshold pressures ($\sigma\kappa_{max}^t$) that are obtained for the throats, $t$, from which the invading phase can fill the pore.    $\kappa_{max}^t$ is the \emph{largest} curvature (positive toward the throat center) encountered by the interface in throat $t$, as it moves toward the pore center.

To obtain $\kappa_{max}^t$, we compute the interface curvature using Eq.\,\ref{eq_pl_k} at two locations and choose the maximum:  (a) when the interface is mid-way between the throat and the pore and (b) when it reaches the pore center.  Note that, except in an (usually) unstable configuration where the phase in the throat is the non-wetting-phase (see Figure \ref{fig_fluid_config}), the maximum curvature is expected to occur at the pore center.  However, the interface can also get pinned (face the maximum curvature) between the throat and the pore center due to the expansion of the throat, specially for contact angles close to 90 degrees where $\cos(\theta+\beta)$ can have its minimum value between the pore and throat centers (see Eq.\,\ref{eq_pl_k} and Figure \ref{fig_Hl_pl}).   



The pore-filling pressures take into account the configuration of wetting fluid in adjacent throats and hence provide an accurate representation of pore filling for complex geometries.  The effect of the fluids in the adjacent throats in Eq.\,\ref{eq_pl_k} is accounted for through the incorporation of layer interface tangent vectors, $\bf s_l$, which depend both on the corner connectivity to corners of its adjacent throats, and on their fluid occupancy; see Appendix \ref{sec_tangents}.  This contrasts with current network models that use empirical formulae to compute pore-filling pressures \citep{0Valvatne2004,0Oren2003}, which may be inaccurate and lead to, for instance, poor predictions of the amount of trapping \citep{0Raeini2015}.

\subsubsection{ Oil layer collapse and growth pressures}

\begin{figure}[H]
\centering 
 \includegraphics[width=0.38\textwidth]{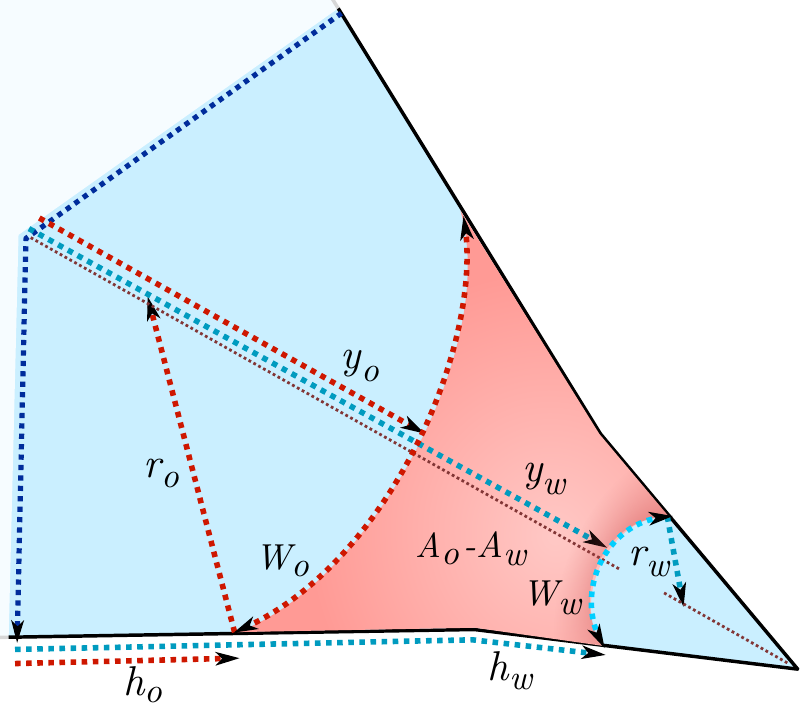}
\caption{An illustration of the parameters used to estimate oil layer collapse/growth pressures. $W$ is the interface length in the corners axial plane. $A_o$ and $A_w$ are the oil and water layer areas. } \label {fig_HlOL} 
\end{figure}

When water invades a throat center, an oil layer can be left behind in the corner if it has a stable configuration -- its collapse pressure is lower than the local capillary pressure.   The oil layer will collapse once the local capillary pressure falls below its threshold collapse pressure. 
The threshold oil layer collapse pressure is obtained using a geometric criterion when it is continuous, connected to oil phase from all its adjacent corners.
It is obtained by iteratively increasing the invading phase (water) pressure until the interfaces on either side of the oil layer join \citep{0Blunt1998}, either from the center:
\be \label{yLayerJoin} y_l^o + y_l^w = 0, \ee
or from the sides:
\be \label{hLayerJoin} h_l^o+h_l^w=0 \ee  
\noindent
where, $h_l^o$ is the location of the interface between the oil layer and the water in the center of the throat  and $h_l^w$ is the location of the water-layer interface, both measured along the sides of the corner;  $y_l^o$ and $y_l^w$ are the locations of the oil and water layer interfaces measured along the center line of the corner,  as shown in Figure \ref{fig_HlOL}.

However, if an oil layer is not continuous from one side  (i.e. it is adjacent to a corner that does not contain oil, neither in its center nor in its crevice) it is expected to collapse more easily.  The entry pressure, ${P_e^{ol}} ={\sigma} \max \kappa_{ol}$ for such scenario is estimated based on a thermodynamic criterion \citep{0vanDijke2007}, by writing a force balance on the interface in the normal direction to the corner's axial plane:
\be \label{eq_thermoLayer}   \kappa_{ol} ({A_o-A_w})  = {2(h_w-h_o)  \cos \theta_o^r } - W_w  - W_o \ee
This equation is solved using the Newton-Raphson method to obtain $\kappa_{ol}$.  

Oil layers can grow in oil-wet corners if the entry pressure required for their growth is lower than the entry pressure for the invasion of the throat center by oil. 
Eq.\,\ref{eq_thermoLayer} is used to obtain the oil layer growth pressure, but with $\theta_o^r$ replaced by $\theta_o^a$.   This thermodynamic criterion is expected to make the layer growth more difficult compared to the geometric criteria (Eqs.\,\ref{yLayerJoin} and \ref{hLayerJoin}).



\subsubsection{ Water layer collapse pressure}

Water layers form if at the maximum depth of the corner ($h_{l,max}$) the condition for their formation is satisfied ($\gamma_3+\theta < \pi /2$) and the center of the throats is filled by the oil phase, but they are assumed not to collapse.

\subsection{ Displacement sequence} \label{sec_filling}

During each flooding cycle, the invading phase pressure at the inlet,  $P_a^{inlet}$, is increased relative to the receding phase pressure, $P_r^{inlet}$.  The local invading phase pressure relative to the receding phase pressures, $P_{a\text{-}r}$, throughout the flow domain is assumed to change proportional to the difference in advancing and receding phase viscous pressures, $\varPhi_{a}-\varPhi_{r}$:
\be P_{a\text{-}r} = P_a^{inlet}-P_r^{inlet}+\varPhi_{a} - \varPhi_{r}.  \ee
where $\varPhi$ is the viscous pressure obtained by solving the mass conservation equation for each phase, discussed in Section \ref{sec_Kr}.
A displacement happens when $P_{a\text{-}r}$ surpasses the entry pressure, $P_e$, for a receding fluid that is adjacent to an invading fluid connected to the inlet.  


Once a fluid is displaced, all the adjacent flow paths that contain the receding phase and are not trapped (are part of a cluster that is connected to the outlet) are considered for subsequent displacement events and their threshold  entry pressures are (re)computed.

Any receding fluid that is part of a cluster not connected to the outlet is considered as trapped. The curvature of trapped fluid interfaces are assumed to be preserved. This implies that the capillary pressures of trapped ganglia remain independent of the imposed capillary pressure at the inlet boundary.      

Furthermore, if an adjacent flow  path contains the invading phase and has been previously marked as part of a trapped ganglion, the ganglion is removed from the trapping list and is brought into capillary equilibrium with the invaded flow path (coalescence event).    
Coalescence events  may require running an  mini-imbibition cycle if the ganglion's local capillary pressure is higher than the system capillary pressure, or a mini-drainage cycle if the ganglion has a lower capillary pressure than the capillary pressure of the invaded flow path.   First, the entry pressures for all the formerly trapped fluids, for filling by the mini-cycle's invading phase, are computed.    In a mini-imbibition cycle, if the entry pressure is higher than the system capillary pressure, the flow paths are filled with the mini-cycle's invading phase.  Once a flow path is invaded, the connectivity of the adjacent receding fluids in the mini-cycle are checked; if they are not connected to the inlet (invasion front), they are marked as trapped -- forming smaller trapped ganglia.

The main displacement cycle is continued by successively increasing $P_{a\text{-}r}$ and displacing the receding fluids that are not trapped and are adjacent to the invading fluids, if their entry pressure is less than $P_{a\text{-}r}$.  This process is continued until the network saturation or capillary pressure reaches a desired user-defined limit.

\subsection{ Computation of relative permeability } \label{sec_Kr}

To compute fluid saturations, permeabilities, and electrical resistivity of the network, we first need to calculate the volume, flow and electrical conductivities of fluids residing in the corners constituting the network.

The layer volume ($V_l$) and electrical conductivity ($g^e_l$) in each corner are assumed to scale linearly with the corner cross-sectional area. We compute them by interpolating the tabulated corner volume and  areas and conductivities, obtained during network extraction:
\be \label {eq_interp_V}
   \varphi_l= \varphi_i+\updelta  \varphi_i\,\frac{A_l-A_i}{\updelta A_i},~~~~ \varphi = V,~ g^e,
\ee
where $A_l$ is given by Eqs.\,\ref{eq_A_l}, \ref{eq_A_h} and \ref{eq_A_CA}. 

To compute the layer flow conductivity in each corner, $g^q_i$, we assume that it scales with cross-sectional area squared:
\be \label {eq_interp_K_q}
g^q_l = g^q_i- \updelta g^q_i\frac{A_l^2-A_i^2}{\updelta  A_i^2},
\ee 
where $\updelta g_i = g_i - g_{i+1}$ and $\updelta  A_i^2=A_i^2-A_{i+1}^2$.  

When there is a piston-like interface separating the fluids in the pore and throat centers, the volume and conductivity of the fluid in the throat are obtained by linear interpolation between the fluid volumes and conductivities of the levels 1 and 2, using the distance of the interface from the throat center, $x_{pl}$, as the interpolation parameter:
\be \label {eq_interp_pl}
   \varphi_{pl}= \varphi_2 +( \varphi_1-\varphi_2)\,{x_{pl}}/{L_{ht}},~~~~ \varphi = V,~ g^q,~ g^e
\ee

The equations above (\ref{eq_interp_V}, \ref{eq_interp_K_q} and \ref{eq_interp_pl}) lead to cumulative fluid volume and conductivities,  the volume and conductivity of all the fluids \emph{below} the interface -- toward the throat surface for piston-like configurations and toward the corner edge in layer configurations.  To compute the area, volume and conductivity of the fluids above the interface -- toward the pore center for piston-like configurations and toward the throat center line in layer configurations -- we simply  subtract the these volumes and conductivities from the single-phase corner volumes and conductivities. The flow conductivities are further multiplied by the new fluid area relative to the area before this subtraction:
\be \label {eq_interp_K_q_centre}
g^q_{centre} = (g^q_{SP}-g^q_l)(A_{SP}-A_l) 
\ee 
where $g^q_{SP}$ and $A_{SP}$ are the level 1 (single-phase) flow conductivity and cross-sectional areas.
 
We compute a single flow rate for each throat and a single pressure for each pore.   The conductivities of the fluids in the corners of each pair of half-throats are averaged to assign a conductivity to each throat.  The averaging is done by grouping the fluids in the corners into two categories, (a) those that are connected together through the fluid occupying the throat center, and (b) those that are not. For the group (a), we first use an arithmetic sum of the  conductivities of fluids in each half-throat, and then take the harmonic average of the two half-throat conductivities.  For the group (b), we first compute the harmonic average conductivity for each corner pairs on opposite sides of the throat surface, to obtain the full-corner conductivity, and then add them together. Finally the two group conductivities are added together to compute the throat conductivities, $g^\alpha_t$, for each phase ($\alpha=o,w$) and for electrical current, $e$.

In addition to the rules above,  for adding the corner conductivities to obtain a single value for the throat conductivity, we assume that layers that are not continuous from both sides do not contribute to the conductivity of the throat. In other words, if a layer is not continuous, including the tailing layers of piston-like configurations, from one side or from either side, its conductivity is not added to the throat conductivity.  Our results, presented in Section \ref{sec_images}, show that this exclusion of discontinuous layer conductivities is essential in predicting the correct behavior of the relative permeability. 

Once the individual throat conductivities are computed, the relative permeabilities of each phase, $\alpha$, are obtained by solving for mass conservation in each pore, $p$, in the network:

\be\label{eq_mat_balance} \sum_{t\in p } q_t^\alpha = \sum_{t\in p } g_t^\alpha(\varPhi_p^\alpha-\varPhi_{nei}^\alpha) = 0, \ee
where $t$ counts for all the throats connected to pore $p$ and $g_t^\alpha$ is the conductivity of the throat connecting the pore to its neighboring pore (subscript $nei$). The summation is done over all the throats connected to the pore.

These equations are solved assuming a dimensionless pressure of 1 at the inlet and 0 at the outlet nodes. The flow rate is then obtained by summing the flow rates entering the flow domain, which is the same as the sum of flow rates in the throats adjacent to the outlet. The relative permeability of each phase is obtained by dividing its flow rate by the single-phase flow rate, which is computing before starting the first cycle using the same boundary condition for pressure. Finally, the computed pressures are scaled to correspond to the flow rate or capillary number assigned for each phase. In the results presented in this paper, for the sake of simplicity, we assume a low capillary number such that the effect of flow rate on the displacement sequence can be ignored. 

The electrical resistivity of the network is obtained using Eq.\,\ref{eq_mat_balance}, but with $g^\alpha$ replaced by $g^e$ and $\varPhi^\alpha$ replaced by the electrical potential, $\varPhi^e$.

\section{ Validation} \label {sec_results}
In the following, we first use analytical approximations and direct simulation of two-phase flow through simple pore-geometries to validate the network model on a pore-by-pore basis.
Then, we compare the generalized network model predictions with a conventional network and core-flood experiments on sandstones. The aim of this comparison is to demonstrate the improvements achieved by the generalized network model in predicting relative permeabilities from micro-CT images of porous rocks.

\subsection{Pore-by-pore validation}

\begin{figure}[H]
\centering 
 \includegraphics[width=0.47\textwidth]{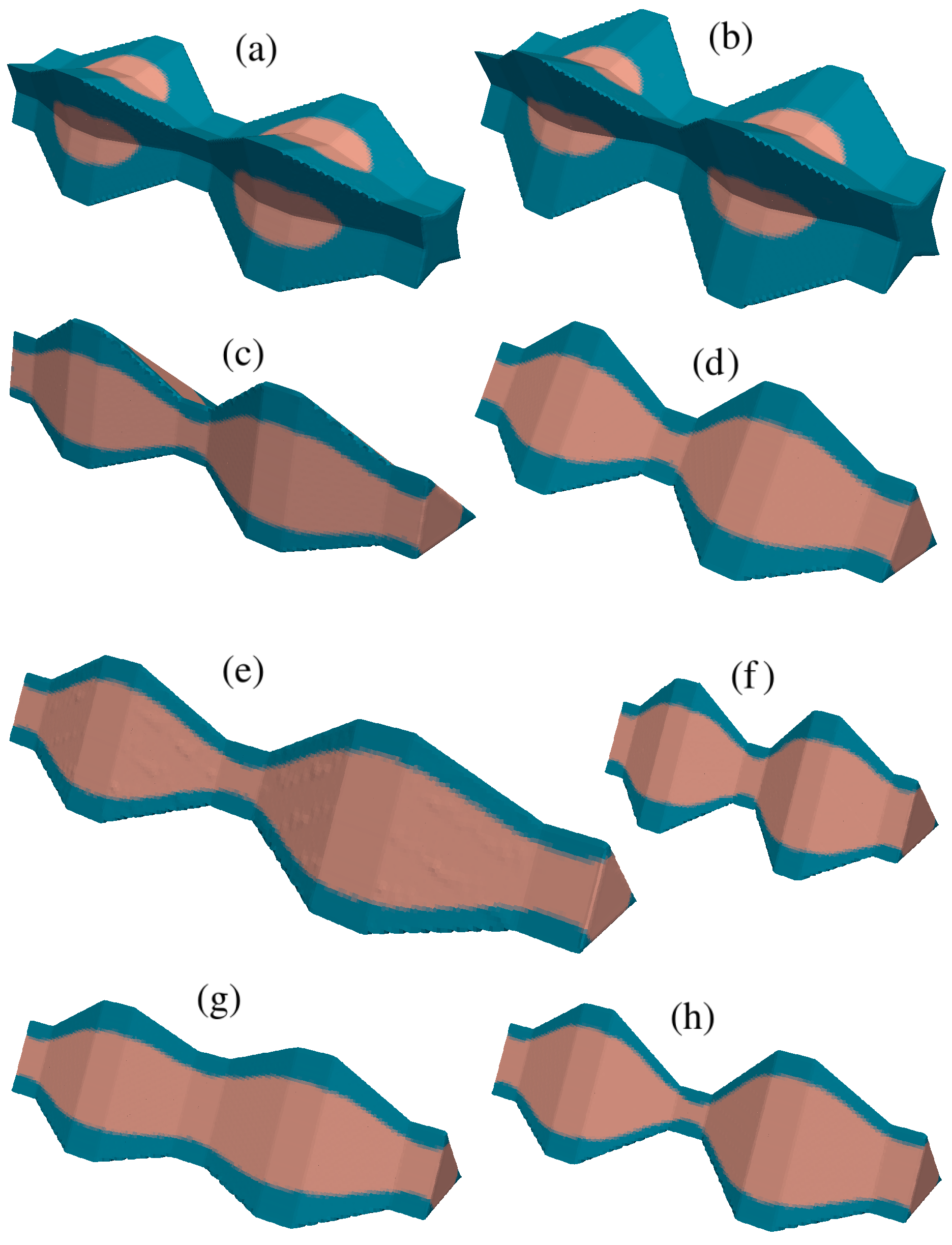} 
 \caption{Synthetic geometries used to validate the network model:  star-shaped geometries with corner angles, $2\gamma$, of (a) 60 and (b) 45 degrees and triangular geometries with corner angles of (c) 40-60-80  and (d-h) 60-60-60 degrees. The aspect ratio ($L_t/R_t$) is 8.33 for all geometries except (e) and (f) for which $L_t/R_t=12.5$ and $6.25$, respectively.  The pore-to-throat contraction ratio ($R_p/R_t$) is 2.5 for all geometries except for (g) and (h) for which $R_p/R_t=1.5$ and 3.5, respectively.  Water is shown in blue and the lighter red color represents the oil phase. The interface is in a piston-like configuration in (a) and (b) and in a layer configuration in the rest.} \label{fig_geomtrySynthetic}
\end{figure}

This section evaluates the accuracy of the network model in calculating the capillary entry pressures, fluid volumes and conductivities on a pore-by-pore basis.  Figure \ref{fig_geomtrySynthetic} shows the synthetic geometries used for this purpose, which include star- and triangular-shaped geometries with different corner angles (described in \citet{0Raeini2017a}), and different pore-throat contraction ($R_p/R_t$) and aspect ($L_t/R_t$) ratios, where $L_t$ is the distance between the two pore centers.   The images are converted to three-dimensional images similar to micro-CT scans with a voxel size of 1.6\unit{\mu m}, corresponding to a resolution of $R_p/\delta x=18.75$, where $\delta x$ is the voxel size, which is also equal to the average grid-block size used in the direct two-phase flow simulations, described below.

\begin{figure}[H]
\centering 
 \includegraphics[width=0.50\textwidth]{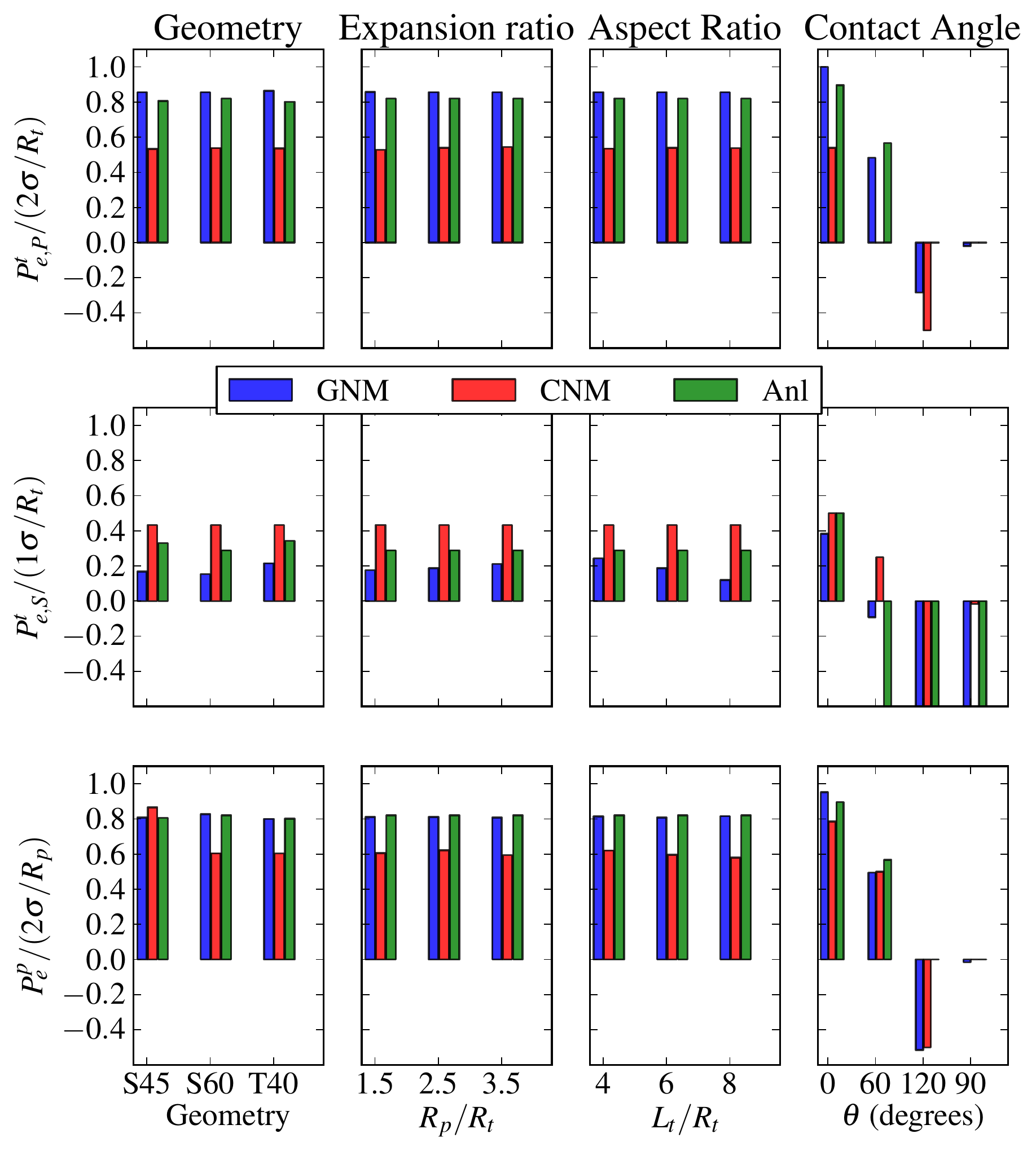}
\caption{Entry pressures for filling the middle throat center by  piston-like (top) and snap-off (middle) invasion, and for pore center filling (bottom). The generalized network model (GNM) is compared to analytical approximations (Anl) and a conventional network model (CNM). } \label{figPes}
\end{figure}

The threshold capillary entry pressures for filling the middle throat by piston-like and snap-off events, and for the pore in the right side of the middle throat are shown in Figure \ref{figPes}.  
The  generalized network model (GNM) results are compared to a conventional network model \citep{0Dong2009,0Valvatne2004} and analytical approximations.  

The analytical approximations (Anl) are obtained using the same equations presented in this paper but using corner angles of the original geometry. However, the effect of interface curvatures in the corner sagittal planes are ignored in their calculation.   

The conventional network model (CNM) uses, in essence, the same equations as the GNM and analytical approximations. However, the number of corners in the CNM and the corner angles are not the same as in the original geometry \citep{0Raeini2017a}.  The interface curvatures in the corner sagittal planes are also ignored in the CNM.


The GNM predictions match the analytical approximations for the entry pressures closely.  The CNM under-predicts the pore entry pressures and over-predicts the snap-off entry pressures, which is because corner angles in the CNM, estimated using shape factors, differ from the original geometry \citep{0Raeini2017a}.  As a result, in the CNM, when the contact angle is 30 degrees the interface in the middle throat at the second cycle snaps off before forming a piston-like configuration and the throat piston-like entry pressures are not computed in the second cycle;  the throat piston-like entry pressures for the CNM shown in Figure \ref{figPes} are taken from the first cycle entry pressures, with the receding contact angles set equal to the advancing contact angles.   
The  GNM predicts a lower snap-off pressure compared to the analytical approximation, which is partly because the sagittal curvature is included in the GNM but not in the analytical approximations.  

To predict relative permeabilities accurately, in addition to entry pressures, the computations of fluid volumes and conductivities should be accurate too.  We validate our computation of fluid volumes and conductivities by comparing the network model predictions with direct two-phase flow simulations on these synthetic geometries, using the same contact angles used in the network model.

The direct numerical simulation (DNS) is a volume-of-fluid based finite volume method. It uses the improved surface tension algorithm  described in \citep{0Shams2017}, and the pressure-velocity coupling and filtering algorithms described in \citep{0Raeini2012}.  An unstructured mesh, with grid blocks roughly the same size as the voxels used in the network extractions (1.6\unit{um}), are used to discretize the flow domain. The grid-blocks away from the solid walls are cubic. However, near the solid walls they are  deformed to align with the solid boundary, using the snappyHexMesh meshing tool from \citet{0OpenFOAM2016}. An additional cell layer is added adjacent to the solid walls, inside the flow domain, so that the wetting layers can be captured more accurately.   The fluid densities and viscosities chosen for both fluids are the same, 1000\unit{kg/m^3} and 0.001\unit{Pa.s}, respectively. The interfacial tension is 0.03\unit{N/m}.  The simulations are performed by first initializing the water layers from the images of corner discretization levels obtained during network extraction. In addition to the three discretization levels discussed in this paper, we run the simulations for a level half-way between discretization levels 1 and 2, called level 1.5, resembling a piston-like configuration; see Figure \ref{fig_geomtrySynthetic}. 

A no-slip boundary condition is used on the solid walls of the direct simulations. The outlet boundary condition is zero-gradient for velocity and indicator function, and a constant value for the dynamic pressure \citep{0Raeini2012}. At the inlet, we have used a zero-gradient boundary condition for all the variables except velocity for which a constant flow rate for each phase is assigned.
The inlet velocities are initially chosen using a zero-gradient boundary condition, but then corrected so that the flow rate of each phase converge toward a desired flow rate \citep{0Raeini2014b}.   The chosen apparent velocities (flow rate divided by image cross-sectional area), for the simulations initialized with images of corner discretization levels 1, 1.5, 2 and 3 are: $q_w=$ 0.6, 0.06, 0.06 and 0.03\unit{mm/s} for the water phase, and  $q_o=$ 0, 0, 0.6 and 0.6\unit{mm/s} for the oil phase, respectively.   Note that oil is not present in the system at level 0, and at level 1.5 it only occupies the pore centers.    These flow rates are sufficiently small that the flow can be considered capillary dominant: the variations in capillary pressure along the middle throat is less than 2\% of the capillary pressure jump across interface, in all of these simulations. 

In summary, four steady-state direct two-phase flow simulations are run for each case, corresponding to four fluid saturations.  Visualizations of the fluid configurations for a set of these simulations, when the flow is steady-state, are shown in Figure \ref{fig_geomtrySynthetic}.  The simulation results are upscaled to compute the saturation and conductivities, for the voxels comprising the middle throat \citep{0Raeini2014b}, and are compared with the generalized and the conventional network model results.

\begin  {figure}[H]
\centering 
 \includegraphics[width=0.49\textwidth]{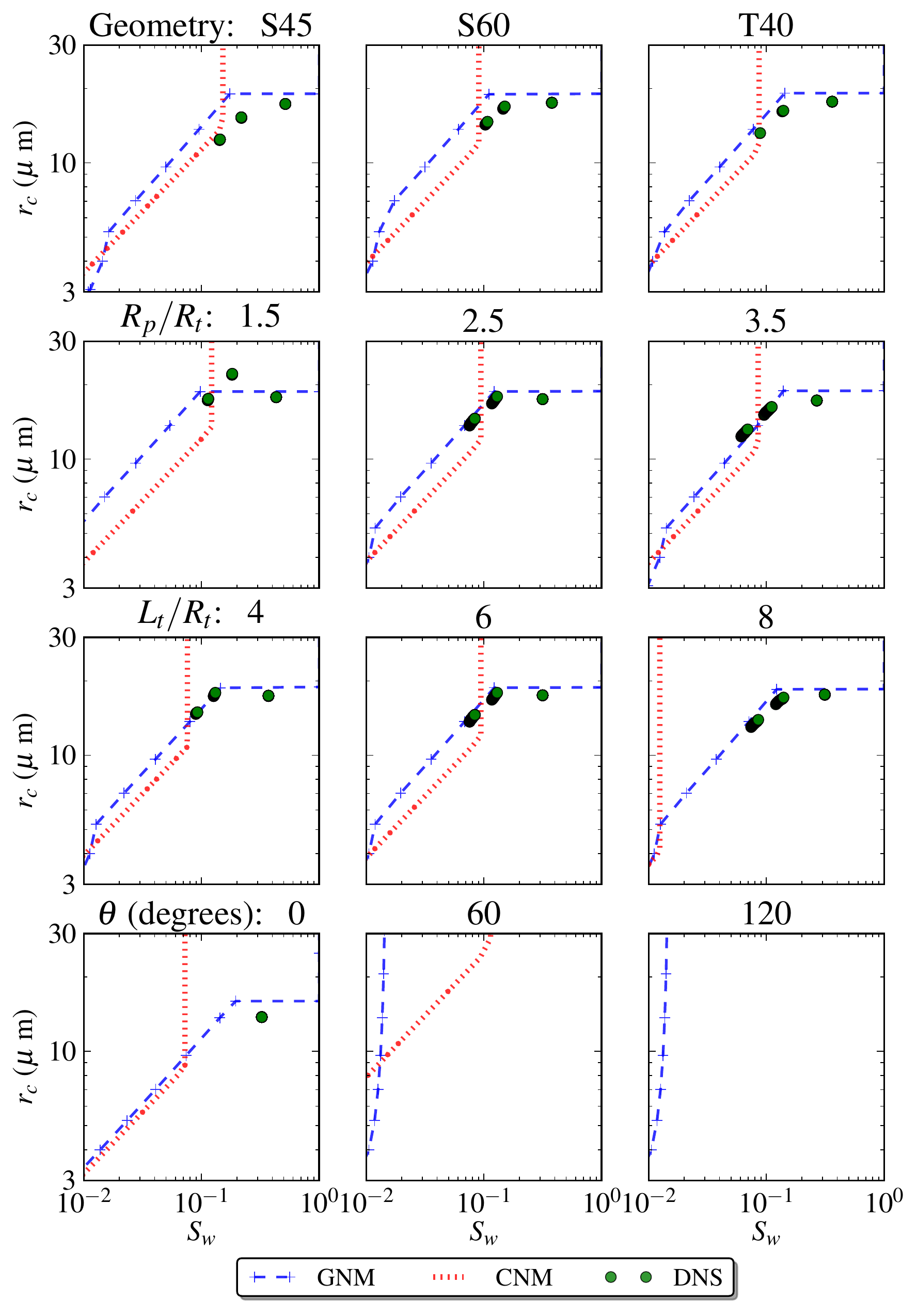}
\caption{Middle throat saturations presented as a function of interface radius of curvature, $r_c=\sigma/P_c$, comparing GNM, CNM and direct two-phase flow simulations (DNS). } \label{figSyntPc}
\end  {figure}

Figure \ref{figSyntPc} shows a comparison between the GNM, CNM and DNS results for the volume fraction of water in the middle throat as a function of curvature radius, as the system capillary pressure decreases during the second (water-flooding) cycle.  

The GNM results match the direct simulations closely, while the CNM over-predicts the corner volumes, for a given curvature radius. Moreover, in the CNM simulations with contact angles of 0 and 30 degrees, all the throats are filled by snap-off, trapping the oil and leading to a fixed saturation as the imposed capillary pressure varies.

Note that the DNS results are only presented for contact angles less than 60 degrees.   To keep the water layer stable for higher contact angles in the DNS,  a hysteresis contact angle, similar to network models should be developed; this is considered a subject for future work. 

\begin  {figure}[H]
\centering 
 \includegraphics[width=0.49\textwidth]{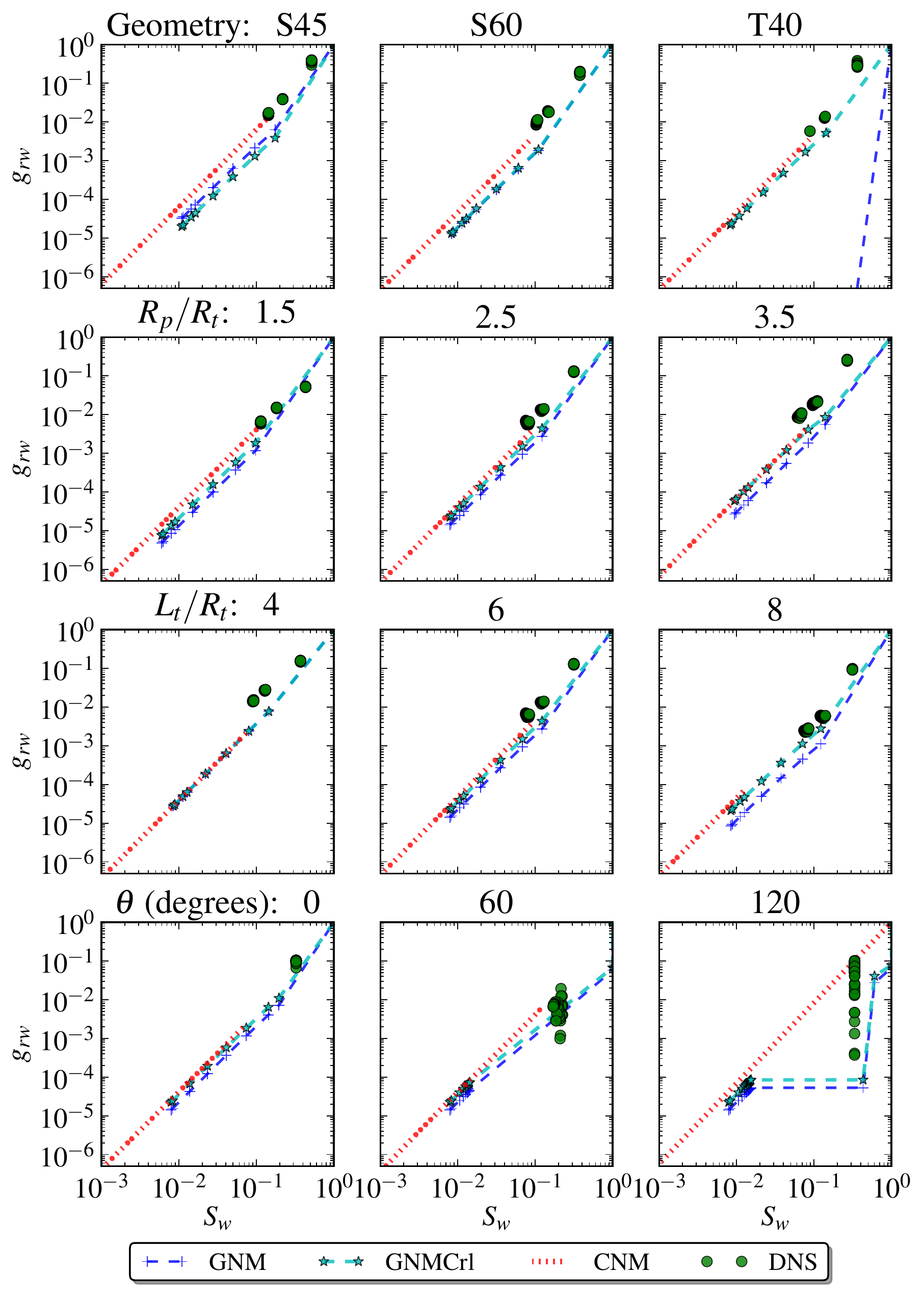}
\caption{Water conductivities normalized by the single-phase flow conductivity, $g_{rw}$, for the middle throat of the synthetic geometries. } \label{figSyntKrw}
\end  {figure}

\begin  {figure}[H]
\centering 
 \includegraphics[width=0.49\textwidth]{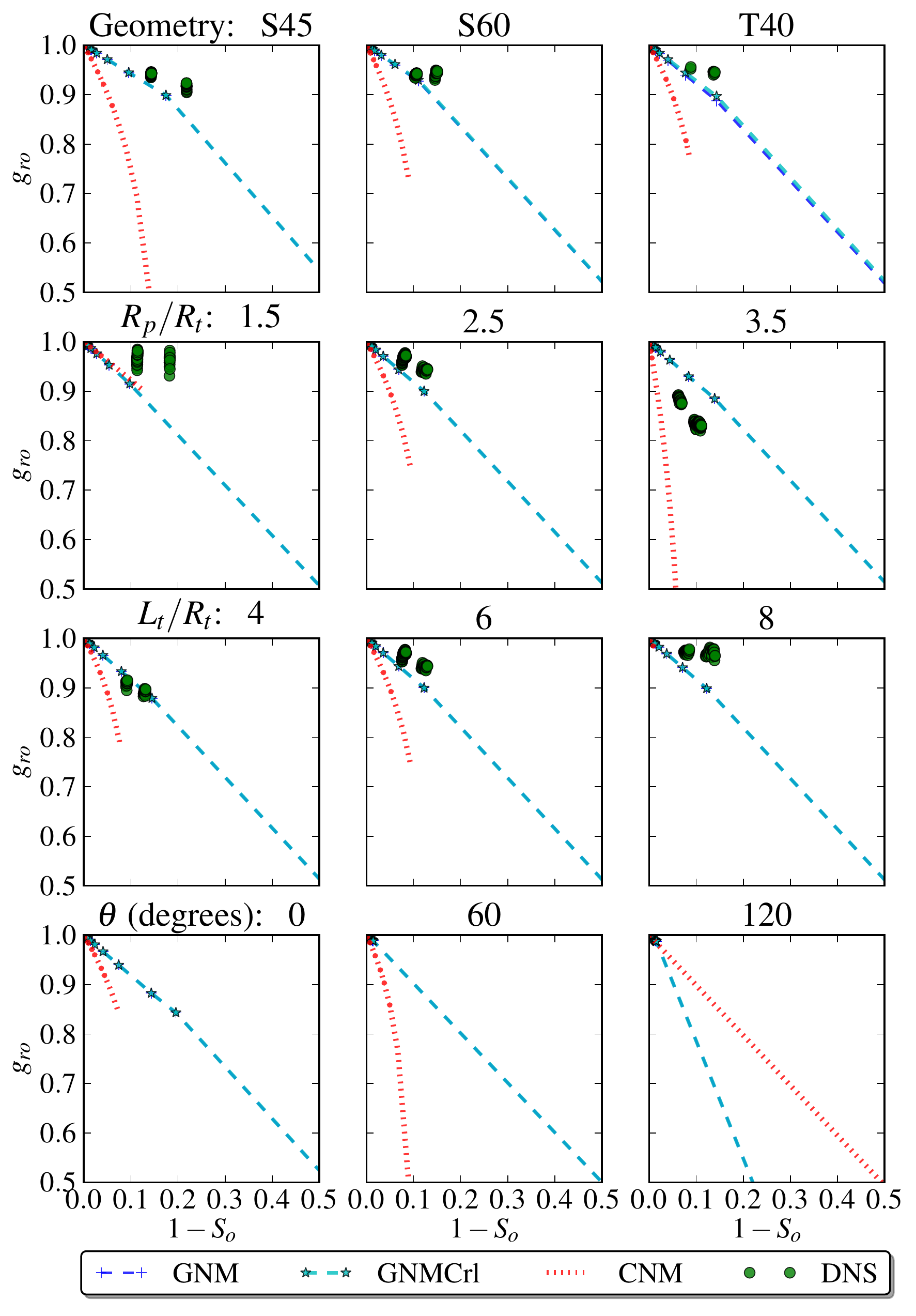}
\caption{Oil phase conductivities normalized by the single-phase flow conductivity, $g_{ro}$, for the middle throat of the synthetic geometries.} \label{figSyntKro}
\end  {figure}

Comparisons between the  GNM, CNM and DNS conductivities for oil and water in the middle throat are given in Figures \ref{figSyntKrw} and \ref{figSyntKro}, respectively. The results, presented for the second (water flooding) cycle, are normalized by single-phase flow conductivities and plotted as a function of saturation.  

We have also compared our results with a generalized network model, but with conductivities obtained using correlations presented in Appendix \ref{sec_condCrl}; this method is called GNMCrl.

The water conductivities in the GNM are underpredicted compared to the direct simulations. This can be explained, 
partly, by the differences in the boundary conditions used in DNS and the direct single-phase simulations used to compute the GNM conductivities.  In the direct single-phase simulations, we have used a slip boundary condition -- no mass and no momentum transfer between the corner voxels and the voxels in the center of throats. The DNS, however, incorporates a continuous velocity across the interface of the two fluids, which implies that there is a drag force between the two fluids, that can increase the apparent conductivity of the water layer \citep{0Raeini2014a}.    Further work is needed to incorporate the effect of viscous coupling and also surface viscosity in the network model \citep{0Dehghanpour2011b,0Xie2017}.

To obtain the GNMCrl results, we originally used correlations \citep{0Raeini2017a} similar to the CNM.  However, the results show a significant over-prediction of the layer conductivities. To fix this problem, we have applied a correction factor, to take into effect the variations in corner angle along the corner, which in turn affects the wetting layer thickness and conductivity, as discussed in Appendix \ref{sec_condCrl}.   Essentially, by applying a correction factor of 0.16, compared to the conductivities used for a corner with a uniform cross section, we have obtained a good match with the DNS results.   This shows the importance of using direct simulations on three-dimensional geometries to compute corner conductivities -- the correlations based on two-dimensional cross-sections of pores and throats are not sufficient.  The use of direct single-phase flow simulation to compute corner conductivities (the GNM formulation) effectively eliminates this source of uncertainty.   

The CNM water conductivities follow the trends of the DNS results.  They do not require the extra correction factor used in the GNMCrl that otherwise uses the same correlations as the CNM.     The choices of pore and throat volumes, lengths and shape factors used in the CNM lead to a good estimation of water conductivities for these geometries and image resolutions. In the following section, however, we show that this statement is not necessarily valid for the more complex case of micro-CT images of porous media; see also \citet{0Raeini2017a} for a discussion on the convergence of CNM and GNM single-phase flow conductivities with image resolution.



\subsection{Micro-CT image of porous rocks} \label{sec_images} 

To assess the predictability of network models of two-phase flow through porous media it is important to consider the uncertainties in the input model description: for example, due to micro-CT image segmentation or variations in the wettability description of the rock sample being studied. The presence of clay and sub-resolution pores should be considered too.  Such a rigorous validation is outside the scope of this work. Nevertheless, in this section we present a set of flow simulations on micro-CT images of porous rocks to demonstrate the improvements achieved by the incorporation of the generalized network model parameters and to show that we can have a reasonable estimation of macroscopic properties of relatively simple rocks with straightforward choices of input parameters. 

We present sample simulations on a Berea and a Bentheimer sandstone, based on $1000^3$ images with voxel sizes of 2.7 and 3.0$\,\mu$m.   The Berea network contains 16595 pores and 36023 throats, while the Bentheimer network contains 8222 pores and 19105 throats.  Other network properties are given in \citep{0Raeini2017a}. During network extraction, the corner images obtained for the discretization level 3 were not connected from the inlet boundary to the outlet, so their conductivity could not be obtained using direct simulations.    Higher resolution images are needed to obtain the level 3 conductivities using direct simulation.  To overcome this problem, we have extrapolated the discretization level 2 conductivities to obtain the conductivities at the level 3: $g_3^q=(R_3/R_2)^4 g_2^q$, and  $\varphi_3=(R_3/R_2)^2 \varphi_2$, where $R_{3/2}=R_3/R_2$ and $\varphi=A,~V,~g^e$. 
This implies that we effectively use two levels to discretize the corners of the micro-CT images.

A uniform intrinsic contact angle of 45 degrees is used in all simulations and Morrow's \citep{0Morrow1975} hysteresis model III is used to compute the receding (oil injection) and advancing (water-injection) contact angles: 3 and 46 degrees, respectively.  

\begin{figure}[H]
\centering 
 \includegraphics[width=0.46\textwidth]{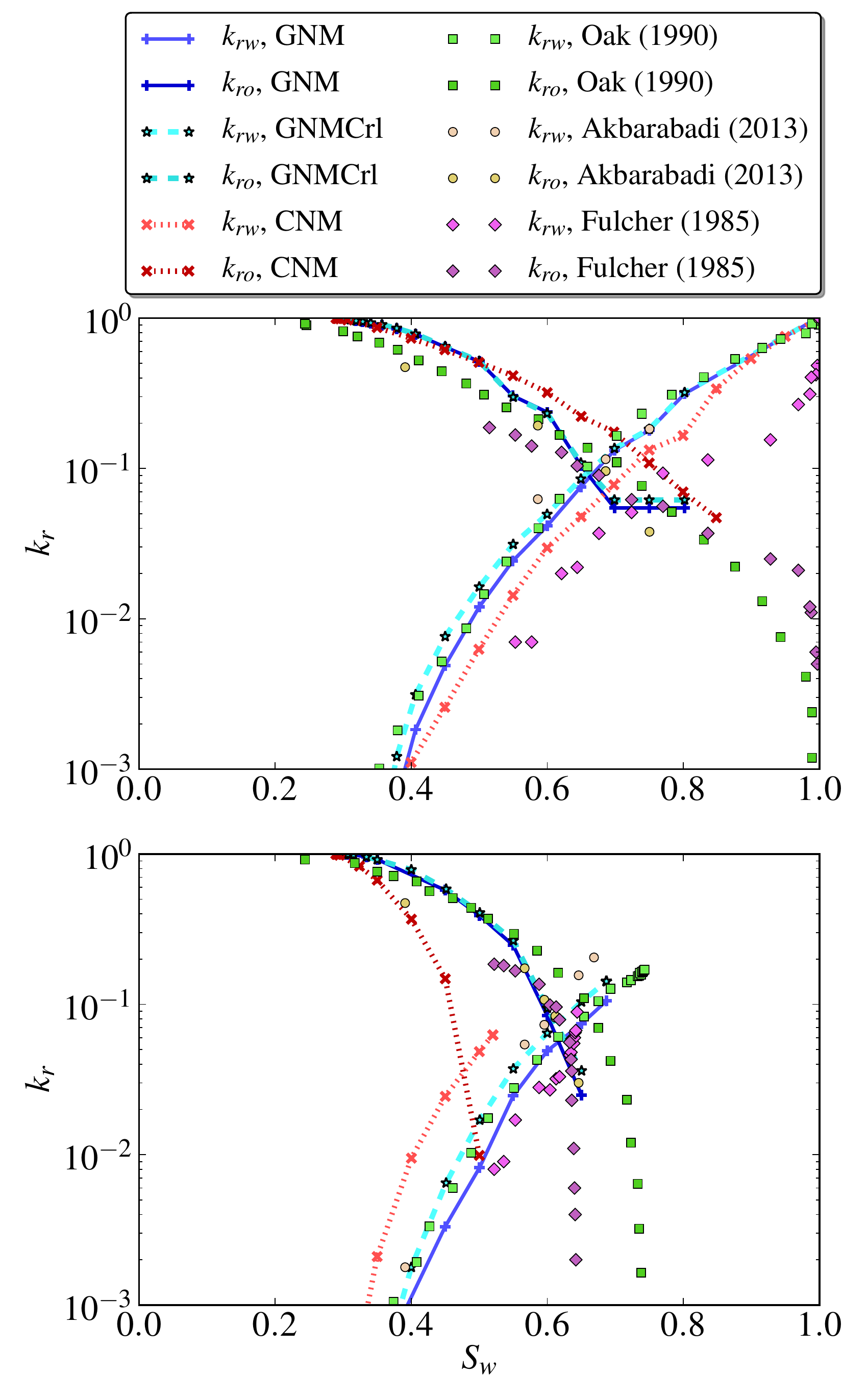} 
 \caption{Drainage (top) and imbibition (bottom) relative permeabilities for Berea sandstone, obtained using different network modeling approaches and compared with the experimental measurements indicated \citep{0Fulcher1985,0Oak1990,0Akbarabadi2013}. 
 A uniform intrinsic contact angle of 45 degrees is used in the simulations. A clay volume of 30\% and 40\% of the pore volume is added to the network, in the GNM and CNM models, respectively.} \label{fig_relPermsBerea} 
\end{figure}

The Berea simulation results are presented in Figure \ref{fig_relPermsBerea}, which are compared to the CNM \citep{0Dong2009,0Valvatne2004} predictions,  and to experimental measurements of relative permeability for oil-water flow by \citet{0Fulcher1985,0Oak1990} and for CO\textsubscript{2}-water flow by \citet{0Akbarabadi2013}.

The GNM relative permeabilities match the Berea experimental data after adding a clay volume of 30\% of the pore volume. This is chosen such that the drainage water relative permeability matches the experimental measurements.  The existence of clay in the Berea is evident from its micro-CT images. The conventional network model, however, requires an higher adjustment of clay volume, 40\% of the pore volume, to match the experimental water relative permeability in the drainage cycle.    Further adjustments to contact angles are needed to obtain a better match for the water relative permeability in the imbibition cycle of the CNM.    Moreover, these results show that the GNMCrl, which uses correlations based on uniform cross-sectional area (two-dimensional geometries), but with a correction factor of 0.16 to the layer conductivity to account for the variations in the corner angle and conductivities along the corner, produces a similar behavior as the GNM formulation.   

\begin{figure}[H]
\centering 
 \includegraphics[width=0.46\textwidth]{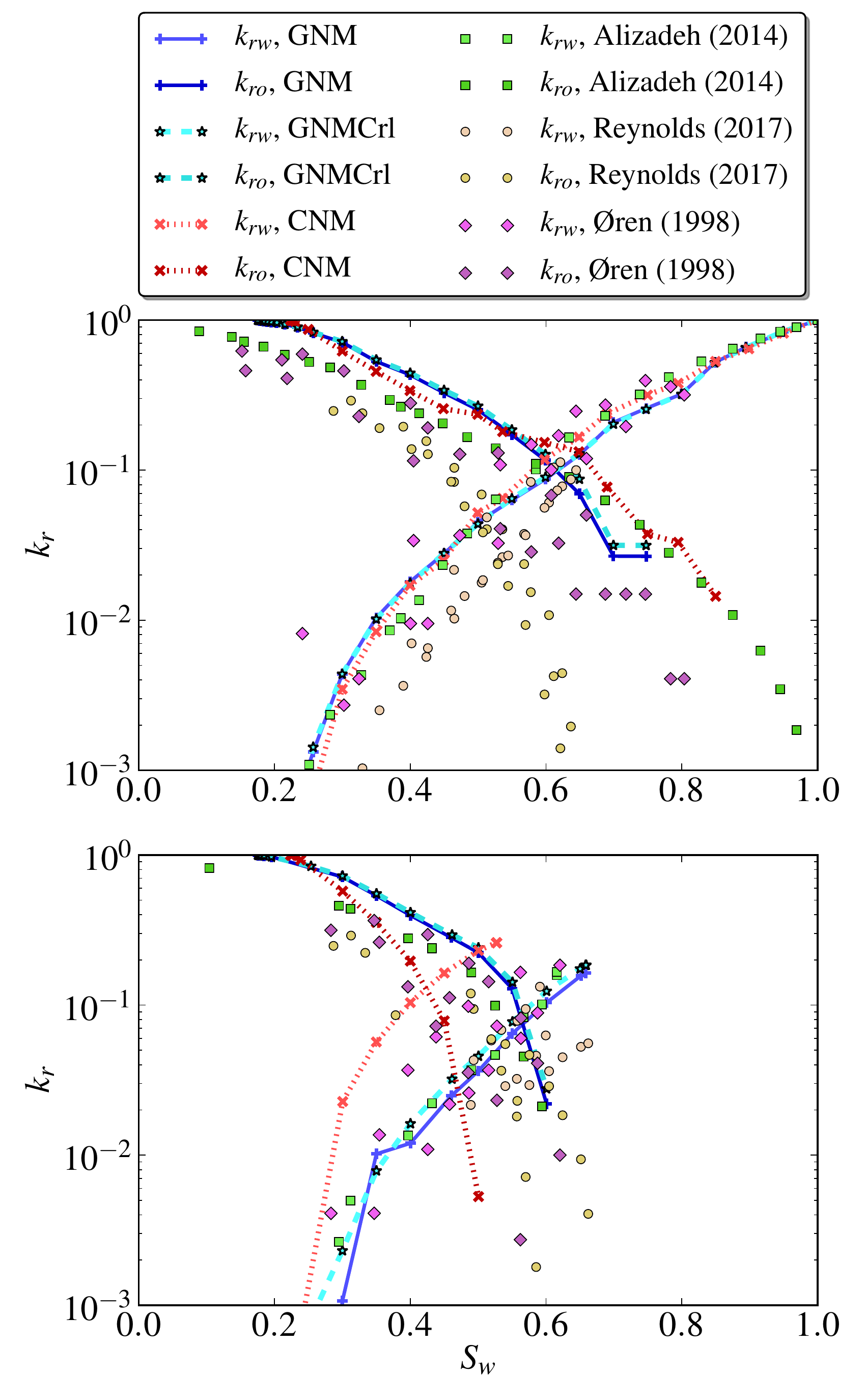} 
 \caption{Drainage (top) and imbibition (bottom) relative permeabilities of Bentheimer sandstone.   The clay volume used, to  match the drainage water relative permeability of the experiments \citep{0Oren1998,0Alizadeh2014a,0Reynolds2017}, is 15\% and 30\% of the pore volume, in the GNM and CNM, respectively.} \label{fig_relPermsBent} 
\end{figure}

The Bentheimer results, Figure \ref{fig_relPermsBent}, are compared to experimental  oil-water relative permeabilities from \citet{0Oren1998, 0Alizadeh2014a}, and to  CO\textsubscript{2}-brine measurements by  \citet{0Reynolds2017,0Krevor2016}.

Both GNM and CNM require a clay volume adjustment to match the experimentally-measured drainage water relative permeabilities of Bentheimer sandstone, 15\% and 30\% of the pore volume, respectively.  This adjustment reflects unresolved porosity, rather than clay itself, as the sandstone has a very low clay content \citep{0Reynolds2014}.  Moreover, the oil relative permeability is slightly over-predicted by the GNM.   The quality of the match can be improved by adjusting throat conductivity and corner connectivity informations so that some water remains almost immobile, acting similar to a clay volume correction -- shifting the relative permeabilities toward higher water saturations -- but also lowering the oil relative permeability.     Further work is needed for a rigorous validation of the generalized network model, with contact angles measured from the micro-CT imaging of two-phase flow \citep{0Andrew2014b,0Aghaei2015} and an assessment of the effect of uncertainties, for instance due to image segmentation and unresolved porosity.

\section{ Conclusions and future work}

We have presented a generalized network model for simulating two-phase flow through micro-CT images of porous media. The network represents a coarse discretization of the pore space with properties obtained from upscaling of direct  simulation of single-phase flow through the corners of the underlying image.

This workflow allowed us to model pore-scale events considering the complexity of the pore geometries encountered in natural porous media and to validate our computations using direct simulation of two-phase flow.
The results show that accurate computations of corner volumes, conductivities and assignment of corner connectivity is critical in the prediction of relative permeabilities from micro-CT images of porous media.

However, there are other sources of uncertainty in the predictions of pore-scale models, related to image resolution, clay volume and corner connectivity, for instance.  To fully resolve these sources of uncertainties,  the network models and validation can be extended by  comparing the results with direct simulations on more complex geometries, considering different wettability distributions and viscous coupling.
Overall, accurate network modeling together with multiphase pore-scale imaging \citep{0Berg2013,0Andrew2015,0Bultreys2015c} and centimeter-scale core-flood experiments offer a predictive framework for linking the pore-scale processes and fluid and rock properties  to their macroscopic properties, helping to answer open questions that cannot be addressed by these methods individually. 

\section{Acknowledgements}
{The authors are grateful to TOTAL for the financial support, fruitful exchanges and permission to publish this work.}

\appendix

\section{Corner shapes and local coordinates}  \label{sec_elemInitAppendix}
\begin{figure}[H]
\centering 
 \includegraphics[width=0.47\textwidth]{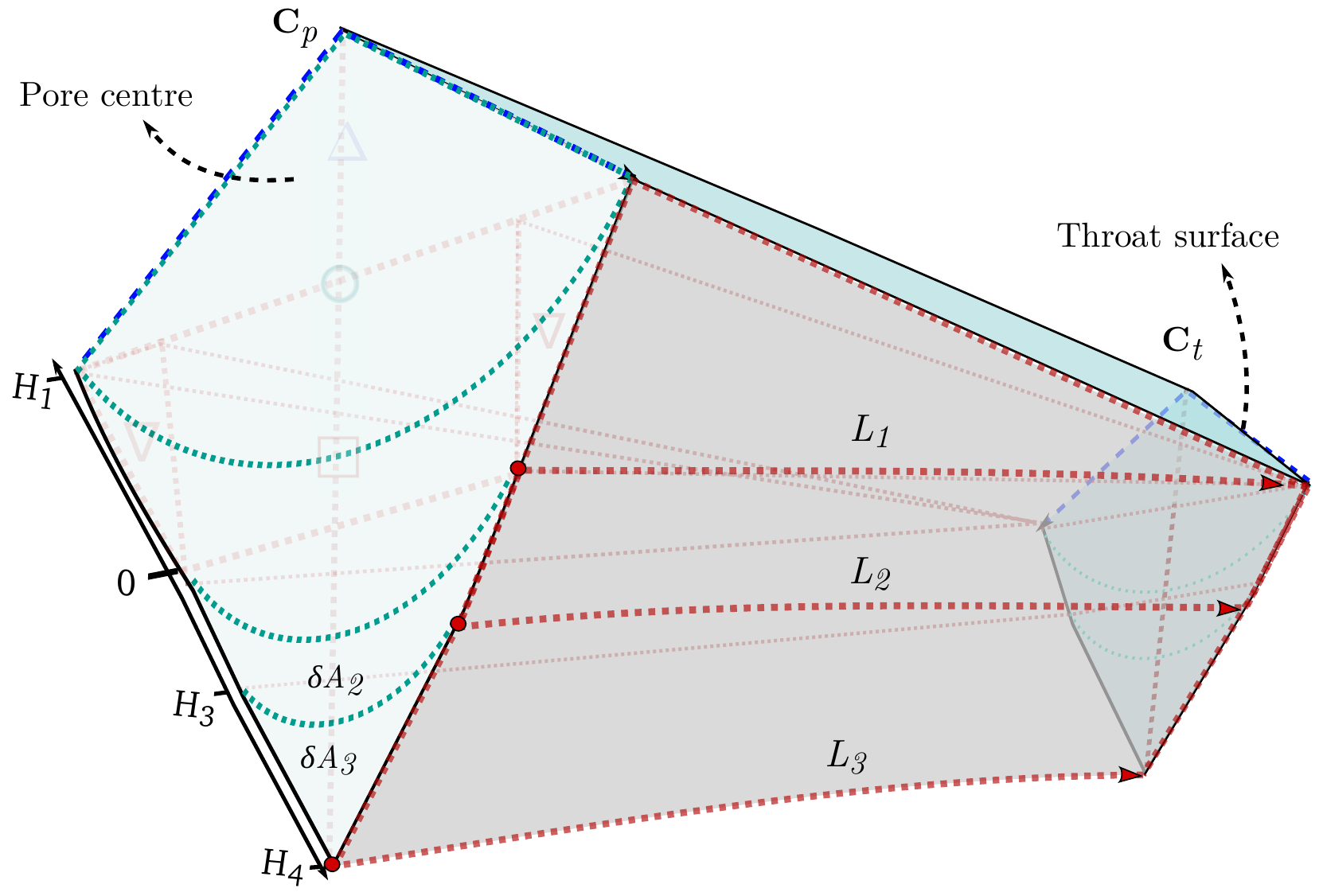} 
 \caption{An illustration of the parameters used to reconstruct the shape of half-throat corners. Note that the cross-sectional areas at levels 2 and 3 are assumed constant along the lengths of the corner. 
 } \label{fig_3dcrnr} 
\end{figure}

The void space in the generalized network model is reconstructed from the parameters extracted during network extraction for each discretization level, $i=1\text{-}3$  as shown in Figures \ref{fig_Hc} and \ref{fig_3dcrnr}.

First, the inscribed radius and cross-sectional areas are used to compute discretization level depths ($H_i$), measured along corner sides and half angles ($\gamma_i$) \citep{0Raeini2017a}:

\be \label{eq_hAng} 
\gamma_i=\sin ^{-1} ( \frac{\cos \gamma_i+\gamma_i\sin \gamma_i}{{\updelta A_i}/{\updelta R_i^2 }+{\frac \pi 2 }} ) \ \ \ i=1\text{-}3
\ee 
\be \label{eq_Hp}
 H_{i+1}  =   H_i+  \frac{\updelta A_i }{\updelta R_i} -{(\frac \pi 2 -\gamma_i)\updelta R_i}, \ \ \ i=1\text{-}3
\ee
where $\updelta$ is the difference operator: $\updelta A=A_i-A_{i+1}$ and $\updelta R=R_i-R_{i+1}$.
$A_1$ and $A_2$ are defined at the throat surface and are assumed not to change between pore and throat centers.  $A_1$ is defined at the pore center.

The discretization level lengths, $L_i$ (see Figure \ref{fig_3dcrnr}) are estimated as follows:

\be \label{eq_L_i}
L_i= \begin{cases}  
 |{\bf C}_t-{\bf C}_p| & i=1 \\
 \\
 \sqrt{L_1^2+(\frac{R_p-R_t}{\sin{\gamma_2}})^2}  & i=2,3 
\end{cases},
\ee

The line connecting throat throat center to the pore center is called the throat axis and is considered as the $x$ axis of the corner. Therefore, the local x coordinate of the pore center is:
\be{x}_p= |{\bf C}_t-{\bf C}_p|\ee

We also compute the vector along the edge of the corner, ${\bf e}_i$ (see Figure \ref{fig_sagital}): 
\be \label{eq_e_i} {\bf e}_i={x}_p{\bf \hat{x}}+\updelta H_1\cos\gamma_1{\bf \hat{y}}_c\ee

The pore distance, $x_p$, and pore and throat radius, $R_i$ are used to compute the angle, $\beta$, between the throat axis and the corner sides (Figure \ref{fig_Hl_pl}).   $\beta$ is computed at the throat center (${x}_t=0$), at the pore center (${x}_p$) and at distance $x={x}_1={x}_p/2$ from the throat surface: 
\be \label{eq_beta}
\beta=
\begin{cases}
 0 & {x}={x}_t,  \\
 \tan^{-1} \dfrac{R_p-R_t}{|{\bf C}_p-{\bf C}_t|} & {x}={x}_1 \\
\cos^{-1}(-\dfrac{{\bf \hat{x}}^j-({\bf \hat{x}}^j\cdot{\bf \hat{y}}){\bf \hat{y}}}{| {\bf \hat{x}}^j-({\bf \hat{x}}^j\cdot{\bf \hat{y}}){\bf \hat{y}} |}\cdot{\bf \hat{x}})   &  {x}={x}_p
\end{cases},
\ee


The total throat cross-sectional area at any point along the throat line is computed as:
\be \label{eq_A_x} A_x^t=\sum A_x^c \ee
where the summation is performed over all the throat corners and $A_x^c$ is the corner cross-sectional area at  a distance $x$ from the throat surface:
\be A_{x}^c =  A_h + (\pi-\gamma_i) R_1^2\ee
where $A_h$ is obtained using Eqs.\,\ref{eq_A_h} at $h=\frac{x_p-x}{x_p}H_1+\frac{x}{x_p}H_2$, located in the discretization level $i=1$.

\section{Interface tangent vectors}  \label{sec_tangents}

The interface tangent vector, ${\bf s}_1$, at $x=x_1$ -- between the pore and throat centers -- is assumed to be parallel to the corner edge vector:
\be \label{eq_s_1} {\bf s}_1 = \frac{ {\bf e}_{\,}  }{ |{\bf e}_{\,}   |}\ee

The interface tangent vectors at the pore and throat centers are also needed to compute the interface curvature in the piston-like configuration.
The interface tangent vector at the throat center, ${\bf s}_t$,  is obtained by averaging the unit edge vectors, ${\bf e}_{~}$ (Eq.\,\ref{eq_e_i}), at either side of the throat surface:
\be
  {\bf s}_t = \frac{ {\bf e}_{~}^{p1}  - {\bf e}_{~}^{p2} }{| {\bf e}_{~}^{p1}  - {\bf e}_{~}^{p2} |}
\ee

For a case that the adjacent corner interface is in a layer configuration, the interface tangent vector at the pore center, ${\bf s}_p$,  is obtained by averaging the unit edge vector with the adjacent corner's unit edge vector:
\be \label{eq_sp_layernei}
  {\bf s}_p = \frac{ {\bf e}_{\,}  - {\bf e}_{\,}^j + ({\bf e}_{\,}^j\cdot{\bf \hat{z}}){\bf \hat{z}} }{| {\bf e}_{\,}  - {\bf e}_{\,}^j +({\bf e}_{\,}^j\cdot{\bf \hat{z}}){\bf \hat{z}} |}
\ee
where ${\bf \hat{z}}$ is the unit vector normal to the throat center line ($x$) and ${\bf \hat{y}}$, pointing away from them.
However, if the adjacent corner's interface is in a piston-like configuration, the interface is expected to bend toward the throat center line. When the piston-like interface is at the pore center, we assume ${\bf s}_p$ is parallel to the throat center line ($x$):
\be \label{eq_sp_pistonnei}
  {\bf s}_p = {\bf \hat{x}}
\ee
Eqs.~\ref{eq_sp_layernei} and \ref{eq_sp_pistonnei} account for cooperative pore body filling \citep{0Lenormand1983, 0Ruspini2017}, which implies that the curvature of the terminal menisci, positive toward the throat center, decreases as more of its surrounding throats are filled with the invading phase,  making filling more favorable..

\section{Correlations for computing corner conductivities} \label{sec_condCrl}

In this section we present a set of correlations that we have used to estimate the corner conductivities in the GNMCrl formulation.  These correlations are considered as a faster but approximate alternative to direct single-phase flow simulations. The effect of viscous coupling and surface viscosity is ignored in these equations.

The corner electrical and flow conductivities (${{g}^e_i}$ and ${{g}^q_i}$, respectively), at discretization levels $i=2$ and $3$, are obtained using the following equations, which are approximations to the correlations used by \citet{0Valvatne2004} for corners of throats with uniform equilateral triangle and square cross-sections.
\be   {{g}^e_i}=\frac{1}{\tau_c}\frac{A_i}{L_i}, ~~~i=2,3 \ee
\be   {{g}^q_i}=\frac{1}{\tau_c^2}(0.168-0.036 \gamma_i) R_i^2 {{g}^e_i}, ~~~i=2,3 \ee
where $\tau_c$ is a correction factor used to account for the changes in the corner angle and cross sectional area along the corner. In this paper, we use $\tau_c=2.5$, which corresponds to a correction factor of $\frac{1}{\tau_c^2}=0.16$ for flow conductivity, chosen such that these correlations reproduce the direct two-phase flow simulations closely.

To estimate the conductivity of corner centers, we first estimate them using the corner parameters at the throat surface:
\be   {{g}^{*e}_1}=\frac{A_1-A_2}{L_1}, \ee
\be   {{g}^{*q}_1}=\frac{R_2^2 {{g}^{*e}_1}}{8-4 A_2/A_1}. \ee
Then, these are corrected for the effect of the expansion of the half-throat cross-sectional area between the throat surface and the adjacent pore centers, assuming that the inscribed radius changes linearly:
\be   {{g}^{**e}_1}  = {{g}^{*e}_1} R_{p/t} \ee
\be   {{g}^{**q}_1} = {{g}^{*q}_1} R_{p/t}^3/(1+ \delta R_{p/t} + \delta R_{p/t}^2/3) \ee
where $R_{p/t} =\frac{R_p}{R_t}$ and $\delta R_{p/t} = \frac{R_p-R_t}{R_t}$ are the expansion ratio and the relative expansion of the inscribed radius from the throat center to the pore center.
Finally, the corner conductivities at level $i=2$ are added to these conductivities to obtain the level 1 (single-phase flow) conductivities.
\be   {{g}^e_1}={{g}^{**e}_1}+{{g}^e_2}, \ee
\be   {{g}^q_1}={{g}^{**q}_1}+{{g}^q_2}. \ee

%


\bibliographystyle{apalike}

\addcontentsline{toc}{chapter}{Bibliography}
\begin{thebibliography}{}

\bibitem[Aghaei and Piri, 2015]{0Aghaei2015}
Aghaei, A. and Piri, M. (2015).
\newblock Direct pore-to-core up-scaling of displacement processes: Dynamic
  pore network modeling and experimentation.
\newblock {\em Journal of Hydrology}, 522:488--509.

\bibitem[Ahrenholz et~al., 2008]{0Ahrenholz2008}
Ahrenholz, B., Tolke, J., Lehmann, P., Peters, A., Kaestner, A., Krafczyk, M.,
  and Durner, W. (2008).
\newblock Prediction of capillary hysteresis in a porous material using
  lattice-{Boltzmann} methods and comparison to experimental data and a
  morphological pore network model.
\newblock {\em Advances in Water Resources}, 31(9):1151--73.

\bibitem[Akbarabadi and Piri, 2013]{0Akbarabadi2013}
Akbarabadi, M. and Piri, M. (2013).
\newblock Relative permeability hysteresis and capillary trapping
  characteristics of supercritical {CO2}/brine systems: An experimental study
  at reservoir conditions.
\newblock {\em Advances in Water Resources}, 52:190--206.

\bibitem[Alizadeh and Piri, 2014]{0Alizadeh2014a}
Alizadeh, A.~H. and Piri, M. (2014).
\newblock The effect of saturation history on three-phase relative
  permeability: An experimental study.
\newblock {\em Water Resources Research}, 50(2):1636--64.

\bibitem[Anderson, 1986]{0Anderson1986b}
Anderson, W.~G. (1986).
\newblock Wettability literature survey -- {P}art 2: Wettability measurement.
\newblock {\em Journal of Petroleum Technology}, 38(11):1246--462.

\bibitem[Andrew et~al., 2014]{0Andrew2014b}
Andrew, M.~G., Bijeljic, B., and Blunt, M.~J. (2014).
\newblock Pore-scale contact angle measurements at reservoir conditions using
  {X}-ray microtomography.
\newblock {\em Advances in Water Resources}, 68:24--31.

\bibitem[Andrew et~al., 2015]{0Andrew2015}
Andrew, M.~G., Menke, H., Blunt, M.~J., and Bijeljic, B. (2015).
\newblock The imaging of dynamic multiphase fluid flow using synchrotron-based
  {X}-ray microtomography at reservoir conditions.
\newblock {\em Transport in Porous Media}, 110(1):1--24.

\bibitem[Arrufat et~al., 2014]{0Arrufat2014}
Arrufat, T., Bondino, I., Zaleski, S., Lagree, B., and Keskes, N. (2014).
\newblock Developments on relative permeability computation in 3{D} rock
  images.
\newblock In {\em Abu Dhabi International Petroleum Exhibition and Conference}.
  SPE-172025-MS.

\bibitem[Berg et~al., 2013]{0Berg2013}
Berg, S., Ott, H., Klapp, S.~A., Schwing, A., Neiteler, R., Brussee, N.,
  Makurat, A., Leu, L., Enzmann, F., Schwarz, J.~O., Kersten, M., Irvine, S.,
  and Stampanoni, M. (2013).
\newblock Real-time {3D} imaging of {Haines} jumps in porous media flow.
\newblock {\em Proceedings of the National Academy of Sciences},
  110(10):3755--59.

\bibitem[Blunt and King, 1991]{0Blunt1991}
Blunt, M. and King, P. (1991).
\newblock Relative permeabilities from two- and three-dimensional pore-scale
  network modelling.
\newblock {\em Transport in Porous Media}, 6(4):407--33.

\bibitem[Blunt, 1998]{0Blunt1998}
Blunt, M.~J. (1998).
\newblock Physically-based network modeling of multiphase flow in
  intermediate-wet porous media.
\newblock {\em Journal of Petroleum Science and Engineering}, 20(3--4):117--25.

\bibitem[Blunt, 2017]{0Blunt2017Book}
Blunt, M.~J. (2017).
\newblock Multiphase flow in permeable media: A pore-scale perspective.
\newblock {\em Cambridge University Press}.

\bibitem[Blunt et~al., 2013]{0Blunt2013}
Blunt, M.~J., Bijeljic, B., Dong, H., Gharbi, O., Iglauer, S., Mostaghimi, P.,
  Paluszny, A., and Pentland, C. (2013).
\newblock Pore-scale imaging and modelling.
\newblock {\em Advances in Water Resources}, 51:197--216.

\bibitem[Blunt et~al., 2002]{0Blunt2002}
Blunt, M.~J., Jackson, M.~D., Piri, M., and Valvatne, P.~H. (2002).
\newblock Detailed physics, predictive capabilities and macroscopic
  consequences for pore-network models of multiphase flow.
\newblock {\em Advances in Water Resources}, 25(8--12):1069--89.

\bibitem[Boek and Venturoli, 2010]{0Boek2010}
Boek, E.~S. and Venturoli, M. (2010).
\newblock Lattice-{Boltzmann} studies of fluid flow in porous media with
  realistic rock geometries.
\newblock {\em Computers {\&} Mathematics with Applications}, 59(7):2305--14.

\bibitem[Bondino et~al., 2013]{0Bondino2013}
Bondino, I., Hamon, G., Kallel, W., and Kachuma, D. (2013).
\newblock Relative permeabilities from simulation in {3D} rock models and
  equivalent pore networks: Critical review and way forward.
\newblock {\em Petrophysics}, 54(6):538--46.

\bibitem[Buckley and Liu, 1998]{0Buckley1998}
Buckley, J.~S. and Liu, Y. (1998).
\newblock Some mechanisms of crude oil/brine/solid interactions.
\newblock {\em Journal of Petroleum Science and Engineering}, 20(3-4):155--60.

\bibitem[Bultreys et~al., 2015]{0Bultreys2015c}
Bultreys, T., Boone, M.~A., Boone, M.~N., De~Schryver, T., Masschaele, B.,
  Van~Loo, D., Van~Hoorebeke, L., and Cnudde, V.~c. (2015).
\newblock Real-time visualization of haines jumps in sandstone with
  laboratory-based microcomputed tomography.
\newblock {\em Real}, 51(10):8668--76.

\bibitem[Dehghanpour et~al., 2011]{0Dehghanpour2011b}
Dehghanpour, H., Aminzadeh, B., and DiCarlo, D.~A. (2011).
\newblock Hydraulic conductance and viscous coupling of three-phase layers in
  angular capillaries.
\newblock {\em Physical Review E}, 83:066320.

\bibitem[Demianov et~al., 2011]{0Demianov2011}
Demianov, A., Dinariev, O., and Evseev, N. (2011).
\newblock Density functional modelling in multiphase compositional
  hydrodynamics.
\newblock {\em The Canadian Journal of Chemical Engineering}, 89(2):206--26.

\bibitem[DiCarlo et~al., 2003]{0DiCarlo2003}
DiCarlo, D.~A., Cidoncha, J. I.~G., and Hickey, C. (2003).
\newblock Acoustic measurements of pore-scale displacements.
\newblock {\em Geophysical Research Letters}, 30.

\bibitem[Dixit et~al., 2000]{0Dixit2000}
Dixit, A.~B., Buckley, J.~S., McDougall, S.~R., and Sorbie, K.~S. (2000).
\newblock Empirical measures of wettability in porous media and the
  relationship between them derived from pore-scale modelling.
\newblock {\em Transport in Porous Media}, 40(1):27--54.

\bibitem[Dong and Blunt, 2009]{0Dong2009}
Dong, H. and Blunt, M.~J. (2009).
\newblock Pore-network extraction from micro-computerized-tomography images.
\newblock {\em Physical Review E}, 80(3):036307.

\bibitem[Dullien, 1992]{0Dullien1992}
Dullien, F. A.~L. (1992).
\newblock {\em Porous media: fluid transport and pore structure}.
\newblock San Diego: Academic Press.

\bibitem[Fatt, 1956]{0Fatt1956}
Fatt, I. (1956).
\newblock The network model of porous media.
\newblock {\em Petroleum Transactions, AIME}, 207:144--181, SPE--574--G.

\bibitem[Fenwick and Blunt, 1998]{0Fenwick1998}
Fenwick, D.~H. and Blunt, M.~J. (1998).
\newblock Three-dimensional modeling of three phase imbibition and drainage.
\newblock {\em Advances in Water Resources}, 21(2):121--43.

\bibitem[Ferrari and Lunati, 2013]{0Ferrari2013}
Ferrari, A. and Lunati, I. (2013).
\newblock Direct numerical simulations of interface dynamics to link capillary
  pressure and total surface energy.
\newblock {\em Advances in Water Resources}, 57:19--31.

\bibitem[Ferréol and Rothman, 1995]{0Ferreol1995}
Ferréol, B. and Rothman, D.~H. (1995).
\newblock Lattice-{Boltzmann} simulations of flow through {Fontainebleau}
  sandstone.
\newblock {\em Transport in Porous Media}, 20(1-2):3--20.

\bibitem[Fischer and Celia, 1999]{0Fischer1999}
Fischer, U. and Celia, M.~A. (1999).
\newblock Prediction of relative and absolute permeabilities for gas and water
  from soil water retention curves using a pore-scale network model.
\newblock {\em Water Resources Research}, 35(4):1089--100.

\bibitem[Fulcher et~al., 1985]{0Fulcher1985}
Fulcher, R., Ertekin, T., and Stahl, C. (1985).
\newblock Effect of capillary number and its constituents on two-phase relative
  permeability curves.
\newblock {\em Journal of Petroleum Technology}, 37(2).

\bibitem[Gostick, 2017]{0Gostick2017}
Gostick, J.~T. (2017).
\newblock Versatile and efficient pore network extraction method using
  marker-based watershed segmentation.
\newblock {\em Physical Review E}.

\bibitem[Gueyffier et~al., 1999]{0Gueyffier1999}
Gueyffier, D., Li, J., Nadim, A., Scardovelli, R., and Zaleski, S. (1999).
\newblock Volume-of-fluid interface tracking with smoothed surface stress
  methods for three-dimensional flows.
\newblock {\em Journal of Computational Physics}, 152(2):423--56.

\bibitem[Hao and Cheng, 2010]{0Hao2010}
Hao, L. and Cheng, P. (2010).
\newblock Pore-scale simulations on relative permeabilities of porous media by
  lattice {Boltzmann} method.
\newblock {\em International Journal of Heat and Mass Transfer},
  53(9--10):1908--13.

\bibitem[Hirt and Nichols, 1981]{0Hirt1981}
Hirt, C.~W. and Nichols, B.~D. (1981).
\newblock Volume of fluid (VOF) method for the dynamics of free boundaries.
\newblock {\em Journal of Computational Physics}, 39(1):201--25.

\bibitem[Huang et~al., 2005]{0Huang2005}
Huang, H., Meakin, P., and Liu, M.~B. (2005).
\newblock Computer simulation of two-phase immiscible fluid motion in
  unsaturated complex fractures using a volume of fluid method.
\newblock {\em Water Resources Research}, 41(12):W12413.

\bibitem[Idowu et~al., 2013]{0Idowu2013}
Idowu, N., Nardi, C., Long, H., {\O}ren, P.~E., and Bondino, I. (2013).
\newblock Improving digital rock physics predictive potential for relative
  permeabilities from equivalent pore networks.
\newblock {\em International Symposium of the Society of Core Analysts, Napa
  Valley, California, 16--19 September, SCA 2013-17}.

\bibitem[Jackson et~al., 2003]{0Jackson2003}
Jackson, M.~D., Valvatne, P.~H., and Blunt, M.~J. (2003).
\newblock Prediction of wettability variation and its impact on flow using
  pore- to reservoir-scale simulations.
\newblock {\em Journal of Petroleum Science and Engineering}, 39(3-4):231--46.

\bibitem[Jadhunandan and Morrow, 1995]{0Jadhunandan1995}
Jadhunandan, P.~P. and Morrow, N.~R. (1995).
\newblock Effect of wettability on waterflood recovery for crude-oil/brine/rock
  systems.
\newblock {\em SPE Reservoir Engineering}, 10(1):40--46.

\bibitem[Koroteev et~al., 2014]{0Koroteev2014}
Koroteev, D., Dinariev, O., Evseev, N., Klemin, D., Nadeev, A., Safonov, S.,
  Gurpinar, O., Berg, S., van Kruijsdijk, C., and Armstrong, R. (2014).
\newblock Direct hydrodynamic simulation of multiphase flow in porous rock.
\newblock {\em Petrophysics}, 55(4):294--303.

\bibitem[Kovscek et~al., 1993]{0Kovscek1993}
Kovscek, A.~R., Wong, H., and Radke, C.~J. (1993).
\newblock A pore-level scenario for the development of mixed wettability in oil
  reservoirs.
\newblock {\em AIChE Journal}, 39(6):1072--85.

\bibitem[Krevor et~al., 2016]{0Krevor2016}
Krevor, S., Reynolds, C., Al-Menhali, A., and Niu, B. (2016).
\newblock The impact of reservoir conditions and rock heterogeneity on co
  2-brine multiphase flow in permeable sandstone.
\newblock {\em Petrophysics}, 57(01):12--18.

\bibitem[Lenormand et~al., 1983]{0Lenormand1983}
Lenormand, R., Zarcone, C., and Sarr, A. (1983).
\newblock Mechanisms of the displacement of one fluid by another in a network
  of capillary ducts.
\newblock {\em Journal of Fluid Mechanics}, 135:337--53.

\bibitem[Li et~al., 2005]{0Li2005}
Li, H., Pan, C., and Miller, C.~T. (2005).
\newblock Pore-scale investigation of viscous coupling effects for two-phase
  flow in porous media.
\newblock {\em Physical Review E}, 72:026705.

\bibitem[L{\o}voll et~al., 2005]{0Lovoll2005}
L{\o}voll, G., M\'eheust, Y., M{\aa}l{\o}y, K.~J., Aker, E., and Schmittbuhl,
  J. (2005).
\newblock Competition of gravity, capillary and viscous forces during drainage
  in a two-dimensional porous medium, a pore scale study.
\newblock {\em Energy}, 30(6):861--72.

\bibitem[Lv et~al., 2017]{0Lv2017}
Lv, P., Liu, Y., Wang, Z., Liu, S., Jiang, L., Chen, J., and Song, Y. (2017).
\newblock In situ local contact angle measurement in a co2--brine--sand system
  using microfocused x-ray ct.
\newblock {\em Langmuir}, 33(14):3358--66.

\bibitem[Man and Jing, 2001]{0Man2001}
Man, H.~N. and Jing, X.~D. (2001).
\newblock Network modeling of strong and intermediate wettability on electrical
  resistivity and capillary pressure.
\newblock {\em Advances in Water Resources}, 24:345--63.

\bibitem[Mani and Mohanty, 1998]{0Mani1998}
Mani, V. and Mohanty, K. (1998).
\newblock Pore-level network modeling of three-phase capillary pressure and
  relative permeability curves.
\newblock {\em SPE Journal}, 3.

\bibitem[Martys and Chen, 1996]{0Martys1996}
Martys, N.~S. and Chen, H. (1996).
\newblock Simulation of multicomponent fluids in complex three-dimensional
  geometries by the lattice {Boltzmann} method.
\newblock {\em Physical Review E}, 53(1):743--50.

\bibitem[Meakin and Tartakovsky, 2009]{0Meakin2009}
Meakin, P. and Tartakovsky, A.~M. (2009).
\newblock Modeling and simulation of pore-scale multiphase fluid flow and
  reactive transport in fractured and porous media.
\newblock {\em Reviews of Geophysics}, 47:RG3002.

\bibitem[Miao et~al., 2017]{0Miao2017}
Miao, X., Gerke, K.~M., and Sizonenko, T.~O. (2017).
\newblock A new way to parameterize hydraulic conductances of pore elements: A
  step towards creating pore-networks without pore shape simplifications.
\newblock {\em Advances in Water Resources}, 105:162--172.

\bibitem[Morrow, 1975]{0Morrow1975}
Morrow, N.~R. (1975).
\newblock Effects of surface roughness on contact angle with special reference
  to petroleum recovery.
\newblock {\em Journal of Canadian Petroleum Technology}, 14:42--53.

\bibitem[Oak and Baker, 1990]{0Oak1990}
Oak, M.~J. and Baker, L.~E. (1990).
\newblock Three-phase relative permeability of {Berea} sandstone.
\newblock {\em Journal of Petroleum Technology}, 42(8):1054--61.

\bibitem[OpenFOAM, 2016]{0OpenFOAM2016}
OpenFOAM (2016).
\newblock The open source cfd toolbox, \url{http://www.openfoam.com}.

\bibitem[Or and Tuller, 1999]{0Or1999}
Or, D. and Tuller, M. (1999).
\newblock Liquid retention and interfacial area in variably saturated porous
  media: Upscaling from single-pore to sample-scale model.
\newblock {\em Water Resources Research}, 35(12):3591--605.

\bibitem[{\O}ren and Bakke, 2003]{0Oren2003}
{\O}ren, P.~E. and Bakke, S. (2003).
\newblock Reconstruction of {Berea} sandstone and pore-scale modelling of
  wettability effects.
\newblock {\em Journal of Petroleum Science and Engineering}, 39(3-4):177--99.

\bibitem[{\O}ren et~al., 1998]{0Oren1998}
{\O}ren, P.~E., Bakke, S., and Arntzen, O.~J. (1998).
\newblock Extending predictive capabilities to network models.
\newblock {\em SPE Journal}, 3(4):324--36.

\bibitem[Pan et~al., 2004]{0Pan2004}
Pan, C., Hilpert, M., and Miller, C.~T. (2004).
\newblock Lattice-{Boltzmann} simulation of two-phase flow in porous media.
\newblock {\em Water Resources Research}, 40(1):W01501.

\bibitem[Patzek, 2001]{0Patzek2001}
Patzek, T.~W. (2001).
\newblock Verification of a complete pore network simulator of drainage and
  imbibition.
\newblock {\em SPE Journal}, 6(2):144--56.

\bibitem[Payatakes, 1982]{0Payatakes1982}
Payatakes, A.~C. (1982).
\newblock Dynamics of oil ganglia during immiscible displacement in water-wet
  porous media.
\newblock {\em Annual Review of Fluid Mechanics}, 14(1):365--93.

\bibitem[Porter et~al., 2009]{0Porter2009}
Porter, M.~L., Schaap, M.~G., and Wildenschild, D. (2009).
\newblock Lattice-{Boltzmann} simulations of the capillary
  pressure-saturation-interfacial area relationship for porous media.
\newblock {\em Advances in Water Resources}, 32(11):1632--40.

\bibitem[Raeini et~al., 2014a]{0Raeini2014b}
Raeini, A.~Q., Bijeljic, B., and Blunt, M.~J. (2014a).
\newblock Direct simulations of two-phase flow on micro-{CT} images of porous
  media and upscaling of pore-scale forces.
\newblock {\em Advances in Water Resources}, 231(17):5653--68.

\bibitem[Raeini et~al., 2014b]{0Raeini2014a}
Raeini, A.~Q., Bijeljic, B., and Blunt, M.~J. (2014b).
\newblock Numerical modelling of subpore scale events in two-phase flow through
  porous media.
\newblock {\em Transport in Porous Media}, 101(2):191--13.

\bibitem[Raeini et~al., 2015]{0Raeini2015}
Raeini, A.~Q., Bijeljic, B., and Blunt, M.~J. (2015).
\newblock Modelling capillary trapping using finite-volume simulation of
  two-phase flow directly on micro-{CT} images.
\newblock {\em Advances in Water Resources}, 83:102--10.

\bibitem[Raeini et~al., 2017]{0Raeini2017a}
Raeini, A.~Q., Bijeljic, B., and Blunt, M.~J. (2017).
\newblock Generalized network modeling: Network extraction as a coarse-scale
  discretization of the void space of porous media.
\newblock {\em Physical Review E}, 96:013312.

\bibitem[Raeini et~al., 2012]{0Raeini2012}
Raeini, A.~Q., Blunt, M.~J., and Bijeljic, B. (2012).
\newblock Modelling two-phase flow in porous media at the pore scale using the
  volume-of-fluid method.
\newblock {\em Journal of Computational Physics}, 231(17):5653--68.

\bibitem[Rajaram et~al., 1997]{0Rajaram1997}
Rajaram, H., Ferrand, L.~A., and Celia, M.~A. (1997).
\newblock Prediction of relative permeabilities for unconsolidated soils using
  pore-scale network models.
\newblock {\em Water Resources Research}, 33(1):43--52.

\bibitem[Ramstad et~al., 2010]{0Ramstad2010}
Ramstad, T., {\O}ren, P.~E., and Bakke, S. (2010).
\newblock Simulation of two-phase flow in reservoir rocks using a lattice
  {Boltzmann} method.
\newblock {\em SPE Journal}, 15(4):917--27.

\bibitem[Regaieg and Moncorg{\'e}, 2017]{0Regaieg2017}
Regaieg, M. and Moncorg{\'e}, A. (2017).
\newblock Adaptive dynamic/quasi-static pore network model for efficient
  multiphase flow simulation.
\newblock {\em Computational Geosciences}, 21(4):1--12.

\bibitem[Reynolds et~al., 2014]{0Reynolds2014}
Reynolds, C., Blunt, M.~J., and Krevor, S. (2014).
\newblock Impact of reservoir conditions on {CO2}-brine relative permeability
  in sandstones.
\newblock {\em Energy Procedia}, 63:5577--85.

\bibitem[Reynolds, 2017]{0Reynolds2017}
Reynolds, C.~A. (2017).
\newblock {\em Two-phase flow behaviour and relative permeability between {CO2}
  and brine in sandstones at the pore and core scales}.
\newblock PhD thesis, Imperial College London.

\bibitem[Ruspini et~al., 2017]{0Ruspini2017}
Ruspini, L.~C., Farokhpoor, R., and {\O}ren, P.~E. (2017).
\newblock Pore-scale modeling of capillary trapping in water-wet porous media:
  A new cooperative pore-body filling model.
\newblock {\em Advances in Water Resources}, 108:1--14.

\bibitem[Sahimi, 1995]{0Sahimi1995}
Sahimi, M. (1995).
\newblock Flow and transport in porous media and fractured rock: from classical
  methods to modern approaches.
\newblock {\em Weinheim: VHC}.

\bibitem[Shams et~al., 2017]{0Shams2017}
Shams, M., Raeini, A.~Q., Blunt, M.~J., and Bijeljic, B. (2017).
\newblock A numerical model of two-phase flow at the micro-scale using the
  volume-of-fluid method framework.
\newblock {\em Manuscript submitted for publication}.

\bibitem[Sheppard et~al., 2005]{0Sheppard2005}
Sheppard, A.~P., Sok, R.~M., and Averdunk, H. (2005).
\newblock Improved pore network extraction methods.
\newblock In {\em International Symposium of the Society of Core Analysts,
  SCA2005-20}, pages 21--25.

\bibitem[Sivanesapillai et~al., 2015]{0Sivanesapillai2015}
Sivanesapillai, R., Falkner, N., Hartmaier, A., and Steeb, H. (2015).
\newblock A {CSF-SPH} method for simulating drainage and imbibition at
  pore-scale resolution while tracking interfacial areas.
\newblock {\em Advances in Water Resources}, 95:212--34.

\bibitem[Tartakovsky et~al., 2009]{0Tartakovsky2009b}
Tartakovsky, A.~M., Meakin, P., and Ward, A.~L. (2009).
\newblock Smoothed particle hydrodynamics model of non-aqueous phase liquid
  flow and dissolution.
\newblock {\em Transport in Porous Media}, 76(1):11--34.

\bibitem[Tsakiroglou and Payatakes, 2000]{0Tsakiroglou2000}
Tsakiroglou, C. and Payatakes, A. (2000).
\newblock Characterization of the pore structure of reservoir rocks with the
  aid of serial sectioning analysis, mercury porosimetry and network
  simulation.
\newblock {\em Advances in Water Resources}, 23(7):773--89.

\bibitem[Tsakiroglou and Fleury, 1999]{0Tsakiroglou1999b}
Tsakiroglou, C.~D. and Fleury, M. (1999).
\newblock Pore network analysis of resistivity index for water-wet porous
  media.
\newblock {\em Transport in Porous Media}, 35(1):89--128.

\bibitem[Tuller et~al., 1999]{0Tuller1999}
Tuller, M., Or, D., and Dudley, L. (1999).
\newblock Adsorption and capillary condensation in porous media: Liquid
  retention and interfacial configurations in angular pores.
\newblock {\em Water Resources Research}, 35(7):1949--64.

\bibitem[Valvatne and Blunt, 2004]{0Valvatne2004}
Valvatne, P.~H. and Blunt, M.~J. (2004).
\newblock Predictive pore-scale modeling of two-phase flow in mixed wet media.
\newblock {\em Water Resources Research}, 40(7):W07406.

\bibitem[van Dijke et~al., 2007]{0vanDijke2007}
van Dijke, M. I.~J., Piri, M., Helland, J.~O., Sorbie, K.~S., Blunt, M.~J., and
  Skj{\ae}veland, S.~M. (2007).
\newblock Criteria for three-fluid configurations including layers in a pore
  with nonuniform wettability.
\newblock {\em Water Resources Research}, 43.

\bibitem[Xie et~al., 2017]{0Xie2017}
Xie, C., Raeini, A.~Q., Wang, Y., Blunt, M.~J., and Wang, M. (2017).
\newblock An improved pore-network model including viscous coupling effects
  using direct simulation by the lattice boltzmann method.
\newblock {\em Advances in Water Resources}, 100:26--34.

\end{thebibliography}

\end{document}